\documentclass[12pt]{iopart}
\usepackage{graphicx,tabularx,color,iopams,cite}

\expandafter\let\csname equation*\endcsname\relax

\expandafter\let\csname endequation*\endcsname\relax

\usepackage{amsmath}

\usepackage[T1]{fontenc}
\usepackage[utf8]{inputenc}

\begin{document}
\title{Bose-Einstein condensates in rotating ring-shaped lattices: a
  multimode model}
\author{M Nigro, P Capuzzi and D M Jezek}
\address{Departamento de Física, Facultad de Ciencias Exactas y
Naturales,  Universidad de Buenos Aires and Instituto de Física
de Buenos Aires (CONICET-UBA), Pabell\'on 1, Ciudad
Universitaria, 1428 Buenos Aires, Argentina } 
\date{\today}

\begin{abstract}
  We develop a multimode model that describes the dynamics on a
  rotating Bose-Einstein condensate confined by a ring-shaped optical
  lattice with large filling numbers. The parameters of the model are
  obtained as a function of the rotation frequency using full 3D
  Gross-Pitaevskii simulations. From such numerical calculations, we
  extract the velocity field induced at each site and analyze the
  relation and the differences between the phase of the hopping
  parameter of our model and the Peierls phase. To this end, a
  detailed discussion of such phases is presented in geometrical terms
  which takes into account the position of the junctions for different
  configurations. For circularly symmetric onsite densities a simple
  analytical relation between the hopping phase and the angular
  momentum is found for arbitrary number of sites. Finally, we
  confront the results of the rotating multimode model dynamics with
  Gross-Pitaevskii simulations finding a perfect agreement.
   
\end{abstract}

\maketitle
\section{INTRODUCTION}\label{introduction}
Over the last decades important efforts have been made to
experimentally investigate the dynamics of Bose-Einstein condensates
(BECs) confined by optical lattices
\cite{Jaksch1998,Greiner2008,Bloch2008,Morsch2006}. Condensates in
different configurations were achieved for several trapping
potentials, including ring-shaped lattices obtained by painting a
time-averaged optical dipole potential on top of a static light sheet
with a rapidly moving laser beam \cite{Henderson}.  At the same time,
plenty of theoretical developments and numerical simulations were
performed in these multiple-well systems (see, e.g.,
\cite{Lewenstein2007,Dutta2015} and references therein).  The
tuning of the optical lattice parameters has also permitted to explore
the quantum phase transition from a Bose-Einstein superfluid phase to
a Mott insulator one \cite{Greiner2002}. However, in the latter case
the increased quantum fluctuations \cite{Altman2005} may invalidate
the theoretical treatment of atomic gases based the Gross-Pitaevskii
(GP) equation \cite{GP}.

For large filling numbers and far from the Mott transition, multimode
models (MM) derived from the GP equation demonstrated to be a useful
and simple tool to predict the evolution of the population and phase
in each site under different scenarios. The accuracy of these models
depends on the adequate calculation of their parameters.  The research
initially addressed two-well systems
\cite{Smerzi1997,Raghavan1999,Ananikian2006,Mele2011,Abad2015,Nigro2017},
where the dynamics can be classified into the Josephson and the
macroscopic quantum self-trapping regimes. These regimes have been
first experimentally confirmed in \cite{Albiez} and implemented by two
weakly linked BECs in a double-well potential. Later on, following the
construction of toroidal traps for the observation of persistent
currents \cite{Ryu2007a}, a laser beam was used to create a single
radial barrier. Such a barrier can act either as a tunable
\cite{Ramanathan2011} or as a rotating \cite{Wright2013} weak
link. The rotating weak link was later used to observe hysteresis in a
quantized superfluid \cite{Eckel2014}.  Two moving barriers have also
been realized using the painting technique to create and manipulate a
BEC in a toroidal trap with a pair of Josephson junctions forming a
double-well system \cite{Ryu2013}. More recently, an experiment in a
ring-shaped optical lattice with $N_c$ tunable barriers was performed
\cite{Aidelsburger2017} where final states with different winding
numbers are formed from up to $N_c=12$ initially uncorrelated
condensates. These experiments provide a promising platform for
studying the nonequilibrium dynamics of atomic gases in ring-shaped
optical lattices. From the theoretical point of view, multimode models
for such ring-shaped configurations have been developed whose
parameters are extracted from the stationary GP states, for either a
double well with two junctions \cite{Cataldo2014}, or an arbitrary
$N_c$ well system \cite{Jezek2013a,Nigro2018a, Nigro2018b}. Such
models have proven to provide very accurate dynamics compared to full
time-dependent GP simulations. The onsite localized functions for
developing the multimode model for $N_c>2$ wells have been constructed
by performing a basis change of the $N_c$ GP stationary states
\cite{pet08} with different winding numbers (or pseudomomentum values)
\cite{Ferrando2005,Perez2007,Cataldo2011,Jezek2011}. Such a basis
transformation \cite{Cataldo2011} can be thought as a generalization
of the superposition of the symmetric and antisymmetric states to
obtain localized functions in double-well systems
\cite{Raghavan1999,Ostrovskaya2000,Mele2011,Abad2015}.  In ring-shaped
lattices the name of Wannier-like (WL) functions has been adopted
\cite{Cataldo2011} in analogy with the so-called localized functions
utilized in solid-state systems \cite{ash76}. However, as discussed in
\cite{Cataldo2011}, it is important to note that the WLs are quite
different in nature to ``true'' Wannier states because in the former
case the occupation number strongly modifies the shape of the
localized functions due to the interaction between particles, as it
also happens in the double-well system. In this context, the onsite
localized functions turn out to be real functions and maximally
localized when all the phases of the $N_c$ stationary states involved
in the basis transformation are fixed equal to zero in the center of a
selected well \cite{Nigro2018a}.  For the construction of the MM
equations \cite{Jezek2013a} a hopping parameter that depends on the
atom interaction has been considered, which was first introduced for a
double-well system in \cite{Ananikian2006}, and an effective 
interaction energy parameter has been also taken into account
\cite{Jezek2013} which has shown to be crucial to correctly describe
the dynamics. Such an effective interaction parameter emerges from the
onsite interaction energy dependence on the population imbalance which
has been disregarded in previous models.

The application of a rotation to the confined systems opened the
possibility to address new matter states and properties of ultracold
atomic gases \cite{Butts1999,Cooper2008,Fetter2009}. First, the
studies were devoted to analyze the superfluid signatures of the
rotating gases in connection with the nucleation and stability of
vortices \cite{Madison2000a,Abo-Shaeer2001,Bretin2004}.  The rotation
of an optical lattice with large filling numbers was utilized in
experiments to observe the vortex nucleation \cite{will10}. In that
work the system was setup in the deep lattice, tight-binding regime
where the depths of the potential wells were such that a 2D array of
weakly linked condensates was created, forming a bosonic Josephson
junction array. The rotation of optical lattices deepened the analogy
to condensed matter physics even further
\cite{Lewenstein2007,Hall,gold14} as the external rotation can be
represented as an additional vector potential with a constant magnetic
field appearing in the rotating frame. The addition of such a vector
potential in turns allows one to build systems with synthetic gauge
potentials \cite{Jaksch2003,Dalibard2011,Aidelsburger2018}, as those
experimentally investigated in
\cite{Leblanc2012,Struck2012,Aidelsburger2013}, realizing the Peierls
substitution for ultracold neutral atoms. In particular, the first
experimental realization of an optical lattice that allowed for the
generation of large tunable homogeneous artificial magnetic fields was
demonstrated in \cite{Aidelsburger2013} with the realization of the
Hofstadter Hamiltonian with ultracold atoms. The studies of synthetic
gauge potentials have also boosted theoretical investigations on
condensates with coupled degrees of freedom providing a renewed
fertile ground for research \cite{Cole2012,Jin2018}.

In this paper we focus on the macroscopic behavior of a BEC confined
by a rotating ring-shaped optical lattice with high filling
numbers. For these trapping potentials velocity fields are induced
inside each well leading to inherently complex WL functions. The goal
of this work is to analyze these imprinted phases and to connect its
behavior with the phase of the hopping parameters arising in a
rotating multimode model (RMM). We pay special attention to circularly
symmetric onsite localized functions for which the prediction of such
phases become simple.  Since the condensates are weakly linked, as a
first step we study the phase profile acquired by single condensates
in off-axis rotating harmonic traps with different aspect ratios by
solving the GP equation. Such findings are also explained by means of
the hydrodynamic equations in the Thomas-Fermi (TF)
approximation. Employing a toroidal trap plus radial barriers it is
shown that the phase profile induced by the rotation in a ring-shaped
optical lattice follows the same behavior as the single condensates.
Moreover, when the induced velocity field in each localized function
is homogeneous the phase of the hopping parameters of the model can be
analytically related to the angular momentum and agrees with the
well-known Peierls phase \cite{Peierls1933}.  However, for
inhomogeneous velocity fields this simple connection is lost.
Finally, we also numerically confirm the RMM model dynamics for
nonstationary states achieving excellent agreement with GP
simulations. For this purpose, we focus on a selected symmetry of the
initial conditions and consider two values of the rotation frequency:
one for which such symmetry is maintained during the whole time
evolution, and another one where this symmetry is not preserved.

The paper is organized as follows. In section \ref{sec:multimode} we
construct the RMM model, define the multimode parameters and derive
the equations of motion. We also explicitly state the confining
potentials considered in this work. In section \ref{impresion} we
first numerically analyze the induced velocity field in off-axis
rotating condensates confined in harmonic traps with different aspect
ratios and provide analytical expressions for the velocity field in
each case. Secondly, we extend these results to ring-shaped optical
lattices. In section \ref{parameters} we investigate the dependence of
the multimode parameters with the rotation frequency and establish the
relation between the phases of hopping parameters and the total
angular momentum when the velocity field is homogeneous.  The energy
spectrum of stationary states and the dynamics of specific states are
studied in sections \ref{stationary} and \ref{nonstationary},
respectively. Finally, in section \ref{summary} we provide a summary
of our work. A discussion on how to select the optical lattice
parameters to obtain uniform velocity fields is included in the
Appendix.
\section{ROTATING MULTIMODE MODEL\label{sec:multimode}}
Rotating traps introduce several new facts when dealing with multimode
models. Phase gradients are induced on the stationary order parameters
inside every well, and hence the localized states cannot be taken as
real functions.  In this section we will first show how to define a
well localized basis set formed by WL functions.  Second, we will
derive the equations of motion including the effective interaction
parameter introduced in \cite{Cataldo2011,Jezek2013,Nigro2018a}.

\subsection{Dynamical equations}
In previous works it has been shown the method for obtaining the
localized states $w_k$ for nonrotating systems
\cite{Nigro2018a,Cataldo2011,Jezek2013a} which are given in terms of
GP stationary states $\psi _n(r,\theta ,z)$, where $n$ label the
corresponding winding number. For large barriers heights
\cite{Jezek2011}, due to the discrete rotational symmetry and charge
inversion processes \cite{Perez-Garcia2007}, the winding number is
restricted to the values $-[(N_c-1)/2]\leq n \leq [N_c/2]$
\cite{Jezek2011}, where $[\cdot]$ denotes the integer part. It has
been shown in \cite{Cataldo2011} that the stationary states with
different winding numbers are orthogonal, and that one can define
orthogonal WL functions localized on each $k$ site given by the
following basis transformation
\begin{equation}
w_k({ r, \theta, z })=\frac{1}{\sqrt{N_c}} \sum_{n} \psi_n({ r, \theta, z})
 \, e^{-i n\theta_k } \,,
\label{wannier} 
\end{equation}
where $\theta _k=2\pi k/N_c$, and $-[(N_c-1)/2]\leq k \leq
[N_c/2]$. It is important to note that the choice of the global phases
of $\psi _n (r,\theta ,z)$ can affect the localization of the WL
functions. A discussion of how to choose such phases in order to
achieve maximum localization is given in \cite{Nigro2018a}. For
nonrotating systems,  (\ref{wannier}) yields real localized WL
functions.

When dealing with rotating optical lattices of $N_c$ wells, we can
also construct an orthonormal basis set $w_k$ with $|w _k|^2$
localized in each $k$ site and defined by (\ref{wannier}), with
the stationary states $\psi_n$ calculated in the rotating frame of
reference. These $N_c$ stationary states thus satisfy
\begin{equation}
\left[ \hat{H}_0 +
g \, N|\psi_n(\mathbf{r})|^2 - {\boldsymbol \Omega}\cdot {\hat{\boldsymbol L}}  \right] \psi_n(\mathbf{r})=\mu _n \psi _n (\mathbf{r}),
\label{GProtstatic}
\end{equation}     
where $\hat{H}_0=-\frac{ \hbar^2}{2m}\nabla^2 + V_{\mathrm{t}}$, being
$V_{\mathrm{t}}$ the trapping potential, and
$\mathbf{\Omega}=\Omega\hat{z}$ is the applied rotation.  Due to the
rotation, the wavefunctions $\psi_n$ have an imprinted velocity field
within each site and carry a spatially inhomogeneous phase
profile. This inhomogeneity in the phase is then transferred to the WL
functions through  (\ref{wannier}).  It is important to mention
that we will remain using an index $n$ restricted to the values
$-[(N_c-1)/2] \le n \le [N_c/2]$ for labeling the stationary
states. In particular, we have imprinted phases to each initial state
with a winding number in such an interval and have obtained the GP
stationary states by means of a numerical minimization of the
energy. As we will see, for $\Omega\neq0$, such a $n$ value could not
necessarily coincide with the actual winding number of the converged
state since it may change in a $N_c$ value during the minimization for
a given $\Omega$. However, since  (\ref{wannier}) is invariant
under the transformation $n\rightarrow n+N_c$ for each $n$ involved in
the summation, both indices could be indistinctly used for obtaining
the localized functions $w_k$.

In the multimode model the order parameter is written employing the WL
basis set as
\begin{equation}
\psi_M ({\mathbf r},t) = \sum_{k} \,  b_k(t)  \,  w_k ({ r, \theta,z})
  \,,
\label{orderparameter}
\end{equation}
with $b_k(t)=\sqrt{n_k(t)}e^{i\phi _k(t)}$. The phase $\phi _k(t)$
does not represent anymore the whole phase in the $k$ site when
$\Omega \neq 0$, but it takes into account its time dependence, while
its spatial profile is carried by the complex WL function $w_k$.
   
The time-dependent GP equation in the rotating frame reads
\begin{equation}
 \left[ \hat{H}_0 +
g \, N|\psi(\mathbf{r},t)|^2 - \Omega \,\hat{L}_z  \right] \psi(\mathbf{r},t)= i\hbar  \, \frac{\partial\psi(\mathbf{r},t)}{\partial \,t} .
\label{gp}
\end{equation}
Inserting the MM model order parameter (\ref{orderparameter}) into
(\ref{gp}), we obtain
\begin{equation}
i\hbar\frac{db_j}{dt}=-\sum _{k}b_k J_{jk}-\sum _{qkl}b_q^* b_k b_l R_{jqkl},
\label{dynb}
\end{equation}
where we have defined
\begin{eqnarray}
&& J_{jk}=-\int d^3r\, w_{j}^*(\hat{H}_0-\Omega \hat{L}_z)w_k, \label{Jjk} \\
&& R_{jqkl}=-gN \int d^3r\, w_{j}^* w_{q}^* w_{k} w_{l}. \label{R} \\
\nonumber
\end{eqnarray}
Since for $\Omega\neq 0$, the localized functions $w_k$ cannot be
assumed as real functions, the parameters (\ref{Jjk}) and (\ref{R})
become complex numbers and deserve a careful analysis.  First of all,
in lattice potentials with high barriers the only relevant values of
$J_{jk}$ and $R_{jqkl}$ involves up to nearest neighbors sites. In
addition, the operator $\hat{H}_0-\Omega\hat{L}_z$ is hermitian so
that the hopping parameter must verify $J_{jk}=J_{kj}^*$. Due to the
discrete symmetry of the ring-shaped lattice potential, only two of
the entire $J_{jk}$ family will be independent, say $J_{00}$ and
$J_{01}$. We define the onsite energy
\begin{equation}
   \epsilon = - J_{kk} = - J _{00} \label{epsilon} 
\end{equation}
and the hopping parameter
\begin{equation}
 J =  J_{kk+1}=J_{01}=-|J|e^{i\theta _J} \label{J} 
\end{equation}
where we have defined the phase associated to $J$, $\theta_J$, so as
to verify $\theta_J=0$ for $\Omega =0$.  We note that for the systems
we shall consider in the following sections, in the nonrotating case,
the computation of $J$ yields a negative value. On the other hand, by
definition we have
\begin{equation}
R_{jqkl}=R_{qjkl}=R_{jqlk}=R_{kljq}^*.
\end{equation}
We exclude terms $R_{j\,j+1\,j\,j+1}$ and $R_{j\,j\,j+1\,j+1}$ which
involve the overlap between the localized densities in neighboring
sites, as these also turn out to be negligible. Then, using again the
discrete symmetry of the trapping potential, there will be only two
independent parameters $R_{jkql}$. Hence, we can define the onsite
interaction parameter $U$ by 
\begin{equation}
  -NU = R_{jjjj}=R_{0000}, \label{U}
\end{equation}
and the interaction-driven hopping parameter $F$ as
\begin{equation}
  R_{0001}=F=|F|e^{i\theta _F} \label{F} .
\end{equation}
Due to the definition  (\ref{wannier}) we also have
$R_{0-1-1-1}=R_{0001}^*$ and
$R_{0111}=R_{000-1}^{*}=R_{0001}$. Inserting all this information and
the definition of $b_{j}$ in  (\ref{dynb}) we finally get the
equations of motion for the populations $n_k$ and phase differences
$\varphi_k=\phi_k-\phi_{k-1}$,
\begin{align}
 \hbar\,\frac{dn_k}{dt} = &  2 |J| \left[ \sqrt{n_k \, n_{k+1}} \, \sin (\varphi_{k+1}+\theta _J) 
- \sqrt{n_k \, n_{k-1} } \, \sin (\varphi_k+\theta _J) \right ]\nonumber\\
-&  2 |F| \left[ \sqrt{n_k \, n_{k+1} } (n_k + n_{k+1} ) \, \sin (\varphi_{k+1}+\theta _F) \right.- 
\left.\sqrt{n_k \, n_{k-1} } (n_k + n_{k-1} ) \, \sin (\varphi_k+\theta _F)\right] \, ,
\label{ncmode1hn}
\end{align}
\begin{align}
  \hbar\,\frac{d\varphi_k}{dt}  = &    ( n_{k-1} -n_{k}) N
  U_{\mathrm{eff}}    
- \alpha (  n_{k-1} - n_{k}) N U \left[   N_c (n_{k-1}+  n_{k})-2 \right]  \nonumber\\
&+  |J| \left[ \left(\sqrt{\frac{n_k}{ n_{k-1}}} 
   - \sqrt{\frac{n_{k-1} }{ n_k}}\,\right) \, \cos (\varphi_k+\theta _J) \right. \nonumber \\
  &+ \left. \sqrt{\frac{n_{k-2}}{ n_{k-1} }} \, \cos (\varphi_{k-1}+\theta _J) 
- \sqrt{\frac{n_{k+1} }{ n_k}} \, \cos (\varphi_{k+1}+\theta _J) \right]\nonumber\\
&-  |F|  \left[ \left( n_k \sqrt{\frac{n_k}{ n_{k-1} }} - n_{k-1} \sqrt{\frac{n_{k-1}}{ n_k}}\,\right)
 \, \cos (\varphi_k+\theta _F)   \right.\nonumber\\
 & + \left( 3\, \sqrt{n_{k-2} \, n_{k-1}} + n_{k-2} \sqrt{\frac{n_{k-2}}{ n_{k-1}}}\,\right) 
     \, \cos (\varphi_{k-1}+\theta _F)
     \nonumber\\
& \left.- \left(  3\, \sqrt{n_{k+1} \, n_k} + n_{k+1} \sqrt{\frac{n_{k+1}} { n_k}}\,\right)  \, 
\cos (\varphi_{k+1}+\theta _F)\right], \nonumber \\
\label{ncmode2hn}
\end{align}
where we have introduced the effective interaction parameter which, as
demonstrated in \cite{Jezek2013a}, consists on replacing $U$ by
$U_{\mathrm{eff}}=(1-\alpha)U$, and including a term proportional to
$\alpha$. The parameter $\alpha$ is determined by the variation of the
onsite interaction energy with the population imbalance
\cite{Jezek2013}. In particular, the onsite interaction energy
parameter decreases when the population on the site increases with
respect to the stationary value because the new, normalized to unity,
onsite density spreads out over a wider region. In the Thomas-Fermi
approximation, $\alpha$ can be exactly calculated and it yields values
of $3/10$, $1/4$ and $1/6$ for 3D, 2D, and 1D systems, respectively
\cite{Jezek2013}. Such an imbalance dependence gives rise to a reduced
effective interaction energy parameter $U_{\text{eff}}$ in the
equations of motion of the model, respect to the commonly used bare
value $U$. The inclusion of $U_{\text{eff}}$ has shown to be crucial
for obtaining a quantitative agreement with the GP calculation in both
2D and 3D multiple well systems \cite{Jezek2013a,Nigro2018a}. The
rotation effects become visible in the equations of motion
(\ref{ncmode1hn})-(\ref{ncmode2hn}) through the complex nature of the
hopping parameters which introduce two shifts $\theta _J$ and
$\theta _F$ in $\varphi _k$.

When arranging a ring-shaped optical lattice with weakly linked
condensates, the imprinted velocity fields on the onsite localized
function will define the values of $\theta_J$ and $\theta_F$. We
anticipate that as the circulation of the velocity field should be
quantized along a closed curve that links the localized functions
$w_k$ through the junctions, the contributions along the site should
be compensated with the phase jumps across the junctions. This means
that when constructing the multimode model such phase jumps should
appear as phases in both the hopping parameters $J$ and $F$. Then,
$\theta_J$ and $\theta_F$ are expected to be equal and thus in general
they will be referred to as
\begin{equation}
\Theta \equiv \theta_J =\theta_F .
\label{eq:THETA}
\end{equation}
The existence of such a phase is consistent with a standard rotating
model where a Peierls phase appears
\cite{Peierls1933,gold14,Aidelsburger2018}. However, we will show that
depending on the shape of the weakly linked condensates the value of
$\Theta$ could not coincide with the usual prediction of the Peierls
phase.
\subsection{\label{sec:trap}Trapping potential}

In our numerical simulations we will consider a BEC of rubidium atoms
confined by two types of trapping potentials which have been
previously experimentally setup \cite{Henderson}. The GP dynamics will
be studied within a four-well ring-shaped trapping potential given by
\begin{equation}
V_1({\bf r} ) =  \frac{ m }{2 }    \left(
\omega_{r}^2  r^2  
+ \omega_{z}^2  z^2 \right)
+   
V_0 \left[  \cos^2(\pi x/q_0)+   
 \cos^2(\pi y/q_0)\right],
\label{eq:trap4}
\end{equation}
where $r^2=x^2+y^2$ and $m$ is the atomic mass. The harmonic
frequencies are given by $ \omega_{r}= 2 \pi \times 70 $ Hz and
$ \omega_{z}= 2 \pi \times 90 $ Hz, and the lattice parameter is
$ q_0= 5.1 \mu$m.  Hereafter, time and energy is given in units of
$\omega_r^{-1}$ and $\hbar\omega_r$, respectively. The coordinates are
given in units of the radial oscillator length
$\ell_r=\sqrt{\hbar/(m\omega_r)}\simeq 1.3\,\mu$m. We also fix the
barrier height parameter at $V_0 = 25 \hbar\omega_r $ and the number
of particles to $ N=10^4 $ to study the dynamics.  On the other hand,
the dependence of the phase impression with the rotation frequency is
also studied for arbitrary number of wells within a lattice potential
given by a toroidal trap with superimposed radial barriers. In
cylindrical coordinates this lattice potential reads
\begin{align}
  V_2(r,z)&=\frac{m}{2}(\omega _r^2 r^2+\omega _z^2 z^2)+V_0 \exp (-2r^2/\lambda _0 ^2)  
  \nonumber \\ &+V_b\sum _k \exp \bigg [-\frac{\left(y\cos\theta _k-x\sin \theta _k\right)^2}{\lambda ^2_b}\bigg ]
  \times\operatorname{H} \left[y\sin \theta _k + x\cos \theta _k\right],
\label{eq:latticeTrap}
\end{align}
where $\operatorname{H}$ denotes the Heaviside function.  The lengths
$\lambda_0$ and $\lambda_b$ are the widths of the central hole and of
the radial barriers, respectively. We fix the parameters
$\lambda_0/\ell_r=3$, $V_0/\hbar\omega_r=50$ and the trapping
frequencies $\omega_r=2\pi\times 70$ Hz and $\omega_z=2\pi\times 90$
Hz.

We will numerically solve the GP equation for both types of potentials
on a grid of up to $512\times 512\times 256$ points and using a
second-order split-step Fourier method for the dynamics with a time
step of $\Delta t = 10^{-4}\omega_r^{-1}$. For more details see
\cite{Nigro2018a}.

\section{The velocity field of a rotating condensate
}\label{impresion}
In subsection \ref{sec:offho} we will first numerically study the
induced velocity field in an off-axis rotating condensate confined by
a harmonic trap. A simple analytical explanation of the velocity field
is given based on the hydrodynamical approach. In subsection \ref{sec:lattoh} we show
that the velocity profiles for the case of the lattice potential
(\ref{eq:latticeTrap}) qualitatively fit into the same general
categories found for off-axis rotating condensates confined by
harmonic traps.
\subsection{\label{sec:offho}Imprinted phases on off-axis rotating
  condensates in harmonic traps}
\subsubsection{Numerical results}
\begin{figure}
\centering
    \includegraphics[height=4cm,clip=true]{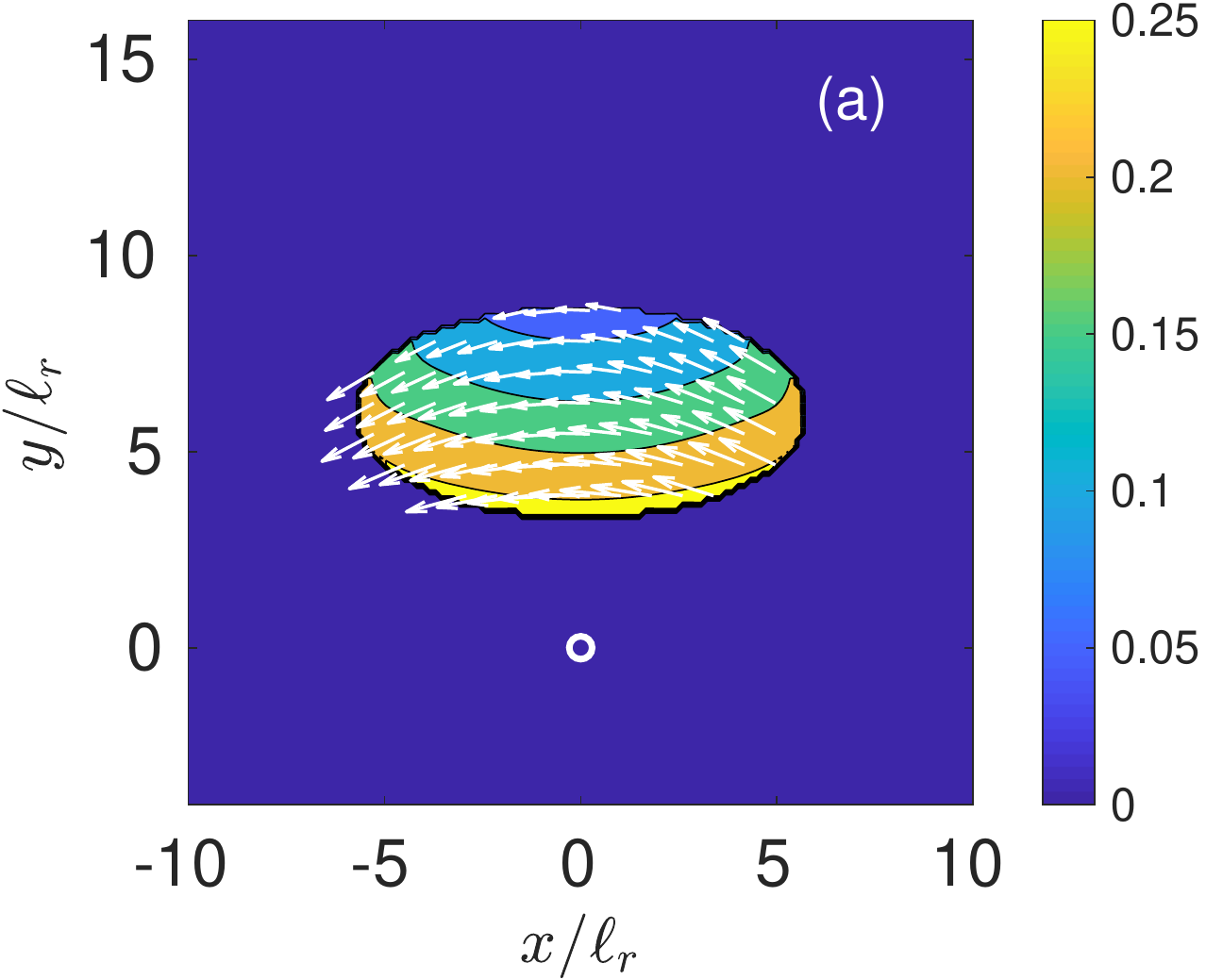} 
    \includegraphics[height=4cm,clip=true]{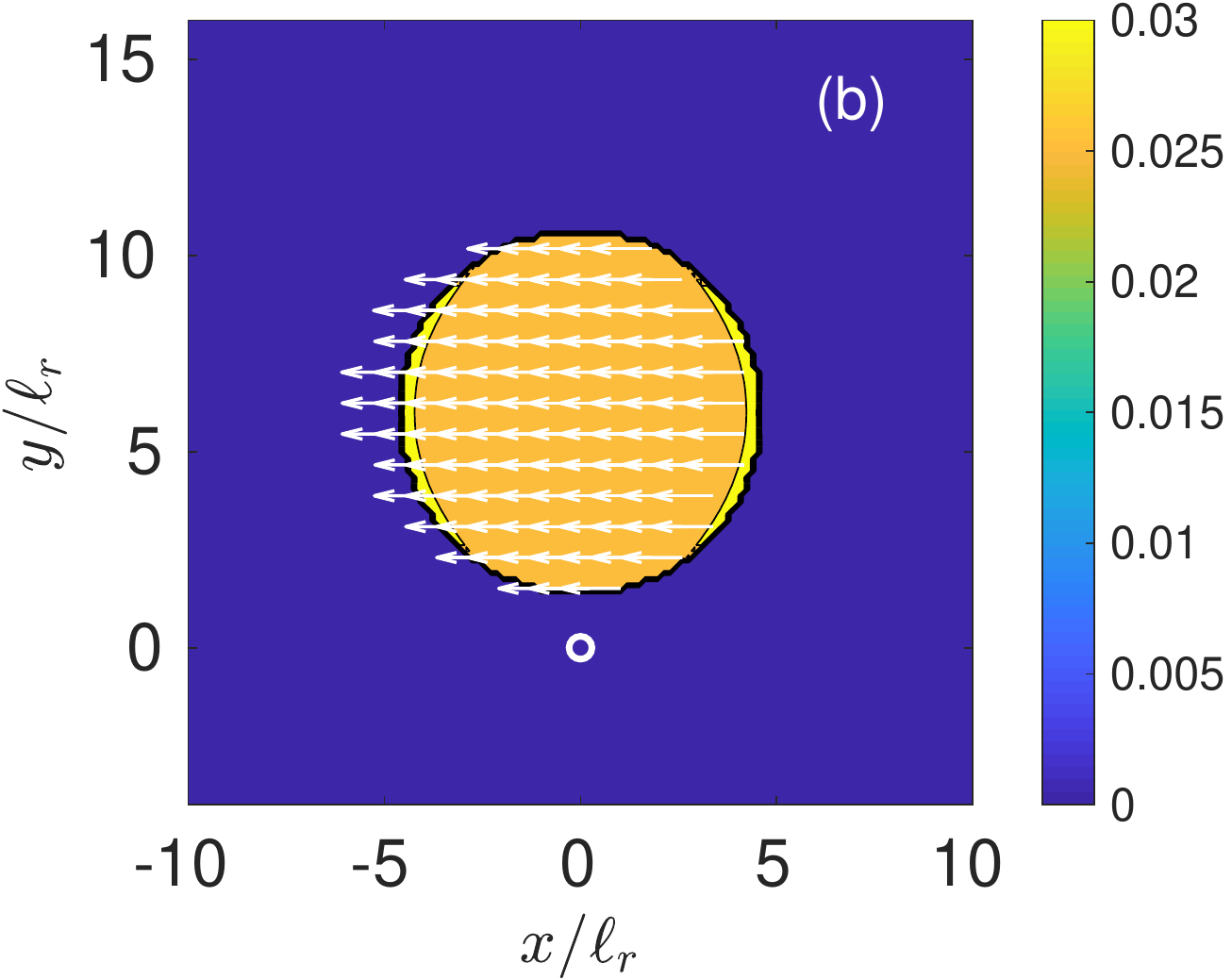} 
    \includegraphics[height=4cm,clip=true]{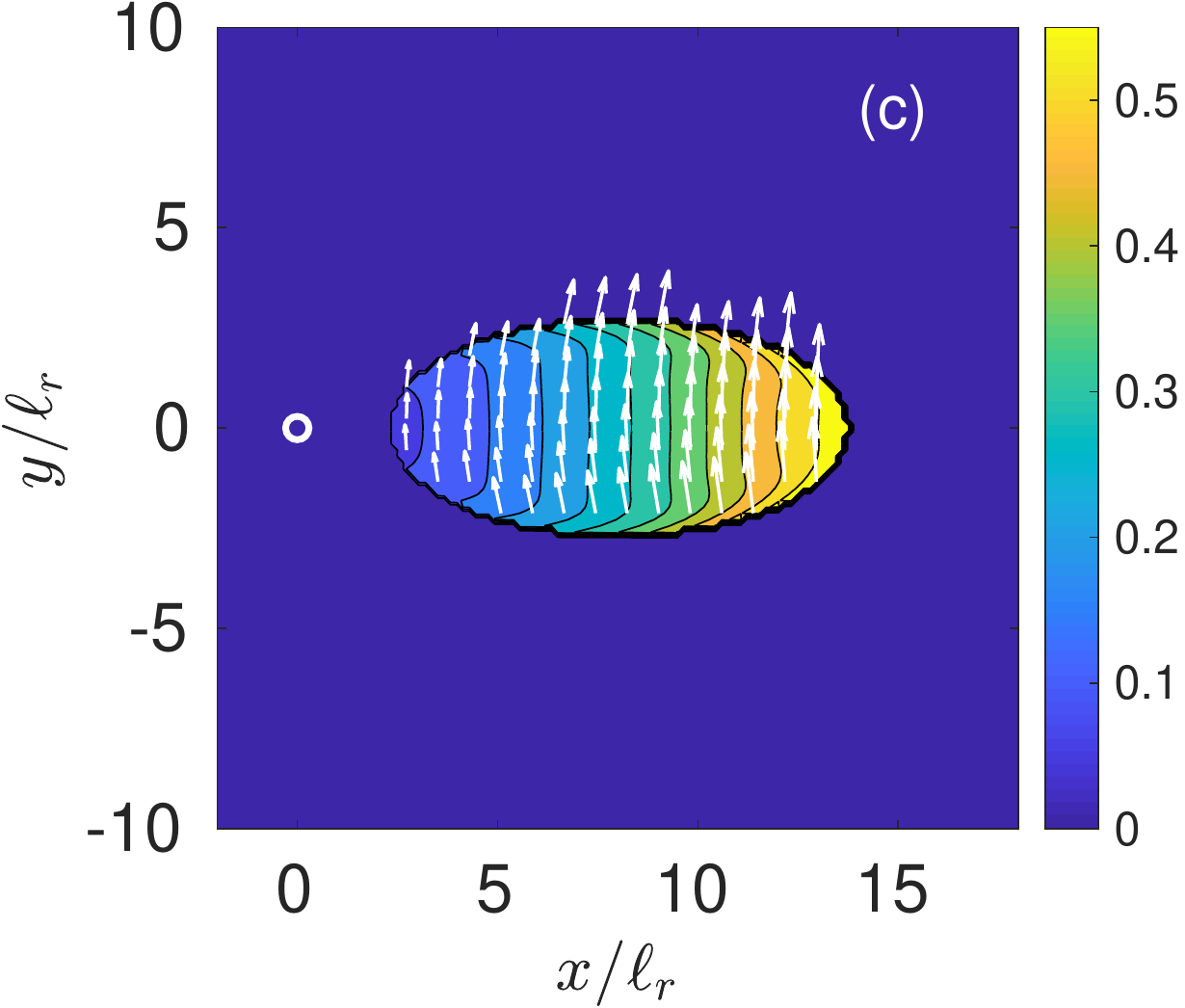} 
  \caption{\label{fig:vs}Velocity fields for off-axis rotating
    condensates with $\Omega=2\pi\times 2$Hz$\simeq 0.03\omega_r$. The
    colors illustrate the squared velocity field contours and the
    arrows represent the velocity field. In each panel the center of
    rotation is depicted with an open circle.}
\end{figure}

When a condensate is subject to rotation the induced velocity field
depends on the geometry of such a condensate and on the location of
the rotation axis. To study the characteristics of such velocity
fields we will consider condensates confined by anisotropic harmonic
traps with different aspect ratios. Previous studies have dealt with
the effects of rotation in centered condensates \cite{Recati2001}. In
this work we focus on condensates whose center is displaced.  We will
vary $ \omega_x $ and fix the other trap frequencies to
$ \omega_y=2\pi\times 70$Hz and
$ \omega_z=2\pi\times 90 $Hz, and the rotation frequency to
$\Omega= 2\pi\times 2$Hz.  The results for the velocity fields are
summarized in figure \ref{fig:vs}, where we depict the velocity vectors
as seen on the laboratory frame together with contours of the squared
velocity modulus. In the top panel we show the velocity field for a
condensate in a trap with $ \omega_x=2\pi\times 30$Hz and whose
center is displaced to $y=6\ell_r$. It may be seen that the velocity
field lines are curved towards the rotation axis following an angular
direction with respect to the rotation axis. We may further see that the
maximum speed is attained at points closest to the rotation axis.

For the isotropic confinement with $ \omega_x=2\pi\times 70$Hz and the
condensate displaced to $y=6\ell_r$, figure \ref{fig:vs}(b), it may be
seen that the field lines are straight and parallel to the $x$-axis,
while their modulus is rather constant. We have also verified that, as
expected from the symmetry, the position of the rotation axis does not
alter the geometry of this induced velocity field.

Finally, in figure \ref{fig:vs}(c) we show a condensate in a trap with
$ \omega_x=2\pi\times 30$Hz and displaced along the $x$-axis to
$x=8\ell_r$. In this case, the velocity field curves outwards respect
to the rotation axis and the maximum speed is reached at the extreme
of the condensate opposite to the rotation axis.
\subsubsection{Hydrodynamical description in the rotating frame}
\begin{figure}
  \centering
  \includegraphics[width=0.8\columnwidth]{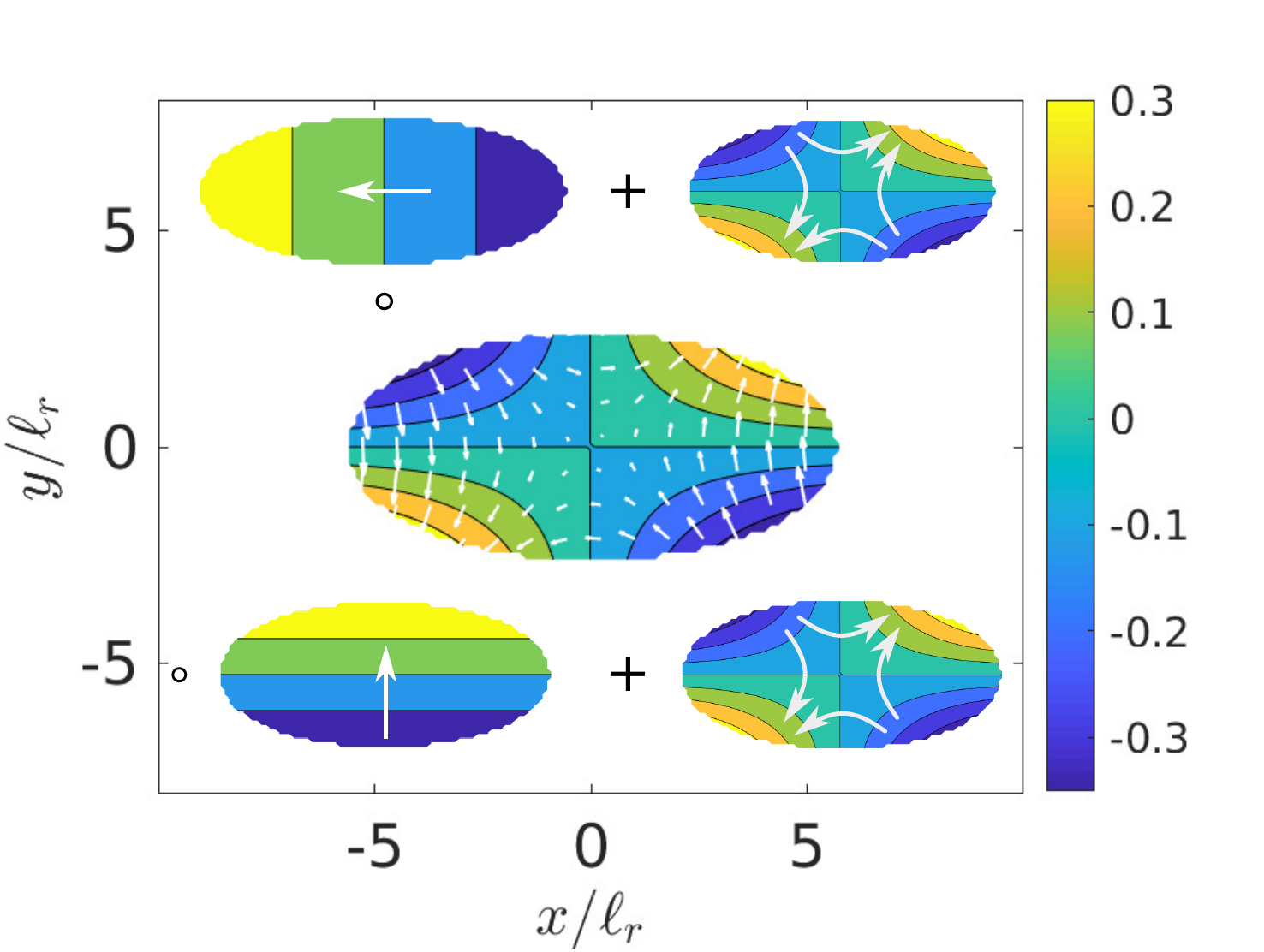}
  \caption{\label{fig:vsesq} The central graph shows the velocity
    field $\delta\mathbf{v}(\mathbf{r})$ together with its phase
    extracted from the GP simulations for the case of the bottom panel
    of figure \ref{fig:vs}. The top and bottom insets show schemes of
    the decomposition of the total velocity field,
    $\mathbf{V}(\mathbf{r})=\mathbf{\Omega}\times\mathbf{r}_{\mathrm{cm}}+\delta\mathbf{v}(\mathbf{r})$,
    where the open circles mark the rotation axes. Such schemes
    represent the cases a) and c) of figure \ref{fig:vs},
    respectively. }
\end{figure}

The GP equation (\ref{gp}) can be written in a hydrodynamical form
following a Madelung transformation $\psi=\sqrt{\rho} e^{i\theta}$,
with $\rho(\mathbf{r},t)$ and $\theta(\mathbf{r},t)$ the density and phase
profiles, respectively. In particular the continuity equation reads,
\begin{equation}
  \frac{\partial \rho( \mathbf{r},t)}{\partial t} + \nabla .  \left[\rho ( \mathbf{V}( \mathbf{r},t)-  \mathbf{ \Omega} \times \mathbf{r})\right] = 0  \,,
\label{continuidad}
\end{equation}
where $\mathbf{V}( \mathbf{r},t)={\hbar}\nabla\theta/m$ is the
superfluid velocity field in the laboratory frame.  The stationary
condition $ {\partial \rho}/{\partial t}=0$ in the rotating frame thus
implies
$\nabla .  [\rho ( \mathbf{V}( \mathbf{r})- \mathbf{ \Omega} \times
\mathbf{r})] = 0 $. We first note that the trivial solution
$\mathbf{V}( \mathbf{r})= \mathbf{ \Omega} \times \mathbf{r}$ is not
irrotational and therefore does not correspond to a superfluid.
However, for an isotropic condensate in the $xy$-plane an homogeneous
velocity field proportional to the center-of-mass position
$\mathbf{r}_{\mathrm{cm}}$ of the form
$\mathbf{V}( \mathbf{r})= \mathbf{ \Omega} \times
\mathbf{r}_{\mathrm{cm}}$ fulfills the above condition given that
$\nabla \rho \perp [ \mathbf{ \Omega} \times
(\mathbf{r}-\mathbf{r}_{\mathrm{cm}} )] $ and
$\nabla . [ \mathbf{ \Omega} \times
(\mathbf{r}-\mathbf{r}_{\mathrm{cm}} )] = 0 $ \cite{note1}.  We note
that such a solution is independent of the particular density profile,
as it only requires that $\nabla \rho$ points in the
$\mathbf{r'}=\mathbf{r}-\mathbf{r}_{\mathrm{cm}}$ direction.  When the
circular symmetry is broken, it is natural to define
$\mathbf{V}( \mathbf{r})= \mathbf{ \Omega} \times
\mathbf{r}_{\mathrm{cm}} + \delta \mathbf{v}( \mathbf{r}) $, where
$ \delta \mathbf{v}( \mathbf{r})$ accounts for the deviation from the
homogeneous value.  For high filling numbers we can resort to the TF
approximation to calculate the density profile in a displaced harmonic
trap. Then, for small $\Omega$ we have
\begin{equation}
\rho \approx \rho _{\mathrm{TF}} =\frac{\mu}{g} -\frac{m}{2g}(\omega ^2_x x'^2 + \omega ^2_y y'^2 + \omega ^2_z z'^2),
\label{TF}
\end{equation}
and (\ref{continuidad}) reads
\begin{equation}
\nabla \rho(\mathbf{r}) \, . \, \delta \mathbf{v}(\mathbf{r})+\rho(\mathbf{r})\, \nabla \cdot \delta \mathbf{v}(\mathbf{r})+\frac{m\Omega}{g} (\omega ^2_y-\omega ^2_x)x'y'=0.
\label{eqpablo2}
\end{equation}
Since the superfluid is irrotational, $\nabla \times \mathbf{V}=0$,
and this implies also that
\begin{equation}
\nabla \times \delta\mathbf{v}(\mathbf{r}) =0.
\label{irrot}
\end{equation}
Given that $\rho$ is a quadratic function of $\mathbf{r}'$, the
solution $\delta \mathbf{v}(\mathbf{r})$ of  (\ref{eqpablo2}) and
(\ref{irrot}) must be linear on the coordinates in the TF
approximation. Moreover, (\ref{irrot}) implies that
$\delta \mathbf{v}=(Ax'+Cy',Cx'+By',0)$.  From  (\ref{eqpablo2}) we
finally obtain
\begin{equation}
\delta \mathbf{v}(\mathbf{r})=\Omega \epsilon _0 (y',x',0),
\label{dv2}
\end{equation}
where
$\epsilon _0 =\frac{\omega ^2_y-\omega ^2_x}{\omega ^2_y+\omega ^2_x}$
measures the anisotropy of the confinement.  The order parameter
can be written as
\begin{equation}
  \psi( \mathbf{r})= | \psi ( \mathbf{r})|  \,   e^{ i
  \frac{m}{\hbar}\left[(\mathbf{r}-\mathbf{r}_{\mathrm{cm}})
  \cdot  (\mathbf{ \Omega} \times \mathbf{r}_{\mathrm{cm}}) +
      \Omega \epsilon _0 (x-x_{\mathrm{cm}})(y-y_{\mathrm{cm}})\right]} ,
\label{impha}
\end{equation}
where we have chosen the phase equal to zero at the position of the
center of mass $\mathbf{r}_{\mathrm{cm}}$.
In figure \ref{fig:vsesq} we show the velocity field
$\delta\mathbf{v}(\mathbf{r})$ extracted from GP simulations and
illustrate how its contribution enhances the squared velocity field in
different regions depending on the location of the rotation axis.
The anisotropy parameter $\epsilon_0$ also enters the angular momentum
of the condensate per particle as
\begin{equation}
  \langle L_z \rangle = m\Omega |\mathbf{r}_{\mathrm{cm}}|^2 +m\Omega \epsilon _0 [(\langle x^2\rangle -x^2_{\mathrm{cm}}) -
  (\langle y^2\rangle -y^2_{\mathrm{cm}})],
\label{Lzgeneral}
\end{equation} 
which in turn shows that $\epsilon_0$ is proportional to the moment of
inertia with respect to the center of mass
$I^{\mathrm{cm}}=\langle L_z^{\mathrm{cm}}\rangle/\Omega =
m\epsilon_0\left(\langle x'^2\rangle - \langle y'^2\rangle\right)$ in
accordance with \cite{Sandro,Recati2001}.  The expression of
$\epsilon_0$ as a function of the trapping frequencies corresponds to
interacting atoms in the TF regime as shown previously by Recati
\textit{et al.} \cite{Recati2001}, whereas the analytic result of
\cite{Sandro} corresponds to a gas with a Gaussian density profile.

When the condensate is circularly symmetric, $\epsilon_0 = 0$, the
velocity field is homogeneous (see figure \ref{fig:vs} (b)), and it
should be equal to
$ \mathbf{v}_{\mathrm{cm}}={\boldsymbol
  \Omega}\times\mathbf{r}_{\mathrm{cm}}$. In such a case the order
parameter takes the simpler form
$\psi( \mathbf{r})= | \psi ( \mathbf{r})| \, e^{ i \frac{m}{\hbar}(\mathbf{r}
  -\mathbf{r}_{\mathrm{cm}})\, . \, (\mathbf{ \Omega} \times
  \mathbf{r}_{\mathrm{cm}})}$. Similarly, for weakly linked condensates
in rotating multiwell confining potentials, the WL function for site
$k$ can be written as
\begin{equation}
  w_k(\mathbf{r}) = |w_k(\mathbf{r})| e^{i \frac{m}{\hbar}(\mathbf{r}-\mathbf{r}_{\mathrm{cm}}^k)\cdot(\mathbf{\Omega}\times\mathbf{r}_{\mathrm{cm}}^k)},
  \label{eq:WLCM}
\end{equation}
where $\mathbf{r}_{\mathrm{cm}}^k$ is the center of mass of the
localized density $|w_k(\mathbf{r})|^2$.

\subsection{\label{sec:lattoh}Imprinted phases in rotating lattices}

In this section we will first consider a four-site rotating lattice
generated by the radial barriers on top of the toroidal trap as given
by (\ref{eq:latticeTrap}) and see how the linked condensates
elongated in the $xy$ plane are transformed into almost circular ones
by varying the values of $V_b$ and $\lambda_b$. This setup permits us
to study the transition of the velocity fields from anisotropic
condensates as that depicted in figure \ref{fig:vs}(a) to circularly
symmetric ones as that shown in figure \ref{fig:vs}(b). Since for the
condensates in this lattice one cannot obtain an analytic solution of
the continuity equation (\ref{continuidad}), we shall directly solve
the GP equation numerically for several lattice parameters.

\begin{figure}
\centering
\begin{tabular}{cc}    
\includegraphics[width=0.4\columnwidth,clip=true]{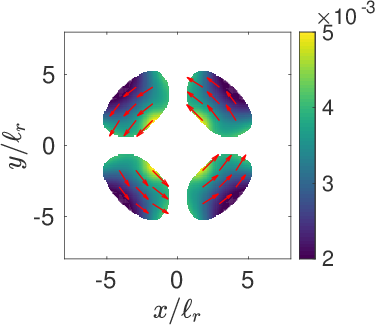}&
\includegraphics[width=0.4\columnwidth,clip=true]{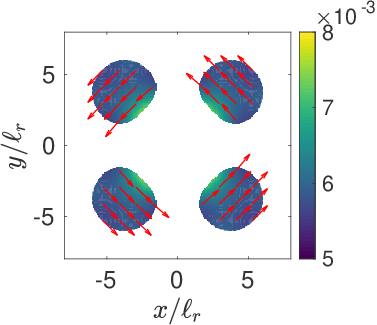}
  \\ (a) & (b) \\
\includegraphics[width=0.4\columnwidth,clip=true]{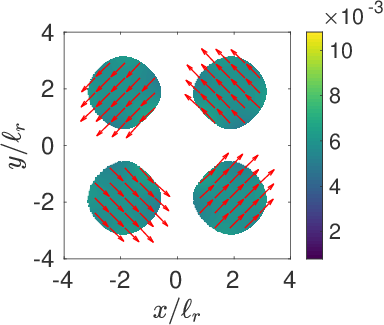}
 & 
   \includegraphics[width=0.4\columnwidth,clip=true]{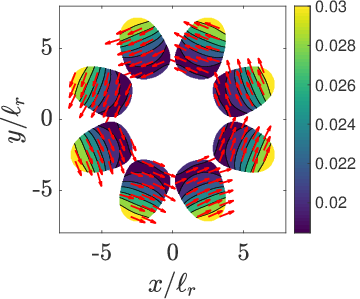} \\
  (c) & (d)
\end{tabular}
\caption{\label{fig:circu} Velocity fields associated to each WL
  function for different rotating traps. The arrows correspond to the
  velocity field calculated with the GP equation and the colors mark
  the squared velocity field value in units of $\hbar^2/(m\ell_r)^2$. In
  (a) and (b) the four-well toroidal trap with radial barriers
  (\ref{eq:latticeTrap}) rotating at $\Omega/(2\pi)=1$Hz was
  considered in order to get inhomogeneous and homogeneous velocity
  fields, respectively fixing $\lambda_b/\ell_r=0.8$,
  $V_b/\hbar\omega_r=15$, and $\lambda_b/\ell_r=3$ and
  $V_b/\hbar\omega_r=21$, respectively. In (c) we employed the
  potential trap given by (\ref{eq:trap4}) and rotating at
  $\Omega/(2\pi)=2$Hz which yields an homogeneous velocity field. In
  (d) we considered eight wells in a lattice potential given by
   (\ref{eq:latticeTrap}) rotating at $\Omega/(2\pi)=2$Hz with
  $\lambda_b/\ell_r=1.6$, and $V_b/\hbar\omega_r=20$, generating a
  velocity profile similar to that in figure \ref{fig:vs} (c).}
\end{figure}

As we showed in section \ref{sec:offho}, the induced velocity field in
an off-axis rotating harmonic condensate acquires a curvature that
tilts in the direction of growth of the velocity field modulus, and
this corresponds to a localized density profile that has no axial
symmetry with respect to its center of mass.  For a lattice in the
tight-binding limit, one expects the same occurs to the induced
velocity fields on the onsite localized WL functions, given that the
effects of the junctions should be negligible.  To observe such a
behavior in ring-shaped optical lattices we numerically obtained the
onsite localized WL function with the potentials introduced in section
\ref{sec:trap} rotating at different angular frequencies $\Omega$. In
figure \ref{fig:circu} we show the imprinted velocity fields, on each
onsite localized function, for distinct trapping potentials: Panel (a)
depicts the results for the lattice potential $V_2$,
(\ref{eq:latticeTrap}), with a narrow radial barrier $\lambda_b$ where
the onsite localized density profiles extend in the angular direction
and the velocity modulus increases when approaching the rotation axis,
similar to figure \ref{fig:vs}(a). In panel (b) each onsite localized
density profile is almost circularly symmetric with respect to its
center of mass and the velocity field is homogeneous, while in figure
\ref{fig:circu}(c) we can observe qualitatively the same profile but
with a potential given by (\ref{eq:trap4}).  Finally, in figure
\ref{fig:circu}(d) we have considered the lattice potential with
$N_c=8$. The parameters $V_b$ and $\lambda_b$ of
(\ref{eq:latticeTrap}) have been chosen in order to obtain onsite
localized functions whose squared velocity field is similar to that
shown figure \ref{fig:vs}(c).  All these findings are in agreement
with the results presented for single condensates subject to off-axis
rotations in harmonic traps. Furthermore, the behavior of the velocity
field curvature can be predicted from the analysis of the balance
between the kinetic energy and the rotation energy terms as briefly
discussed in the Appendix.

\section{THE MULTIMODE PARAMETERS} \label{parameters}

\subsection{Modulus of the parameters}
\begin{figure}
\centering
  \includegraphics[width=0.7\columnwidth]{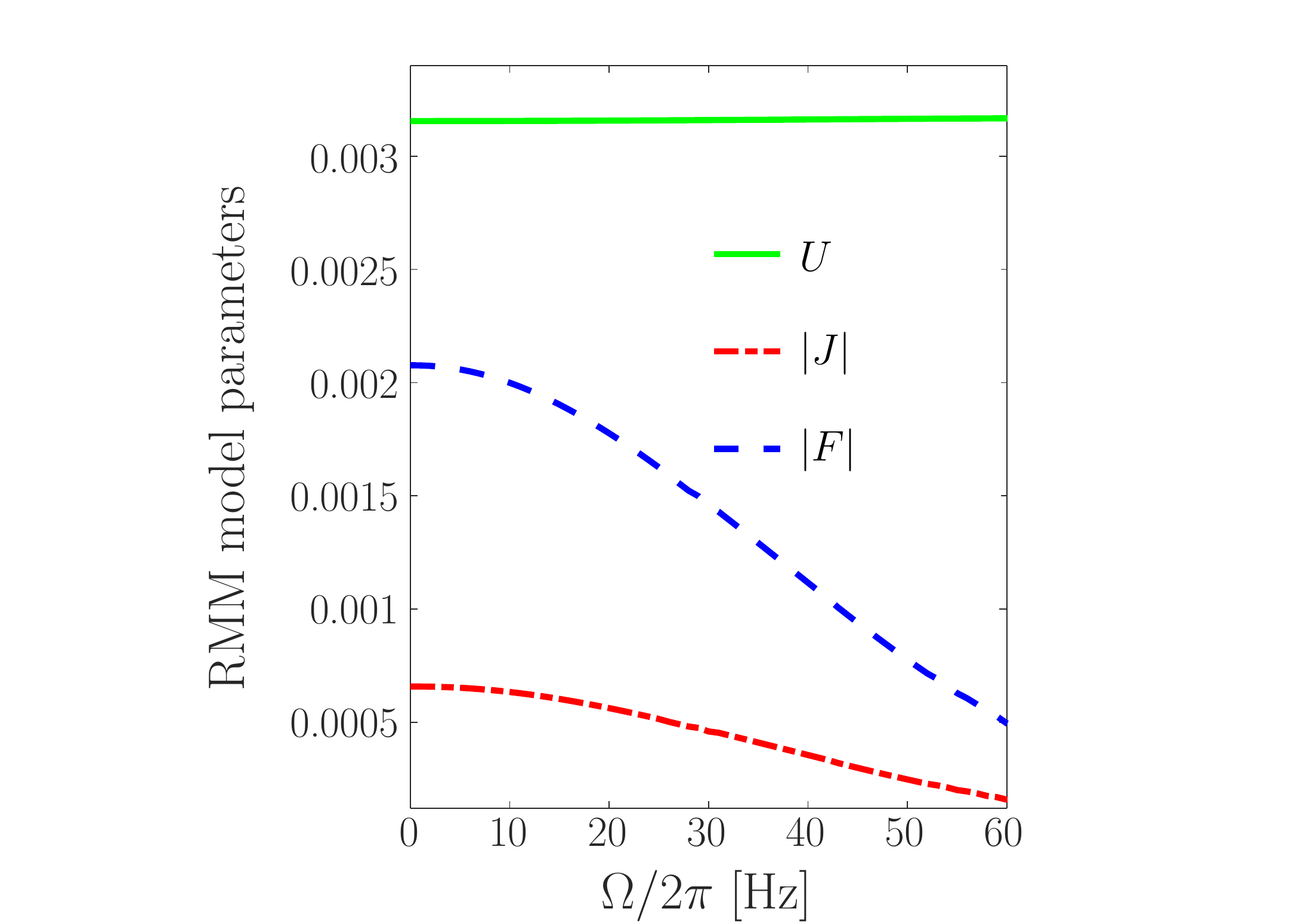}
  \caption{\label{fig:UJF} Absolute values of the RMM model parameters
    (in units of $\hbar\omega_r$) as functions of the rotation
    frequency $\Omega$ for the four-site potential well 
    (\ref{eq:trap4}). }
\end{figure}

In a rotating lattice the localized WL functions are modified respect
to the nonrotating case due to the effective centrifugal force that
opposes to the harmonic confinement. We thus expect all the model
parameters to be affected: the onsite interaction parameter $U$ in our
case is likely to increase, while the modulus of the hopping
parameters $|J|$ and $|F|$ are expected to decrease as the density
moves away from the center leading to a smaller overlap between
neighboring WL functions.  We have numerically investigated the model
parameters for a condensate confined by the potential (\ref{eq:trap4})
as a function of $\Omega$.  The rotation frequency has been varied
keeping $\Omega < \omega _r$ to ensure the equilibrium of the
condensate \cite{Recati2001,Cooper2008}. In figure \ref{fig:UJF} we
summarize the results.
As it can be seen in the figure, the effect of the rotation on the
onsite interaction parameter $U$ is negligible, while $|J|$ and $|F|$
strongly decrease as $\Omega$ gets larger. Therefore, the Peierls
substitution $J\rightarrow | J| e^{i\Theta}$
($F\rightarrow |F| e^{i\Theta}$) comprising only a change in the phase
of the hopping parameters does not suffice as the modulus of $J$ and
$F$ also depend on $\Omega$.

\subsection{The phase of the hopping parameters}
\subsubsection{Relation between $\Theta$ and the  velocity field circulation}
The complex nature of the hopping parameters introduces the shift
$\Theta$ given by (\ref{eq:THETA}). We will see that one can also
determine $\Theta$ by using the order parameter of the MM model
$\psi_M$ of (\ref{orderparameter}) to calculate the velocity field
circulation along a closed curve that passes through each junction.

\begin{figure}
\centering
  \includegraphics[width=0.7\columnwidth]{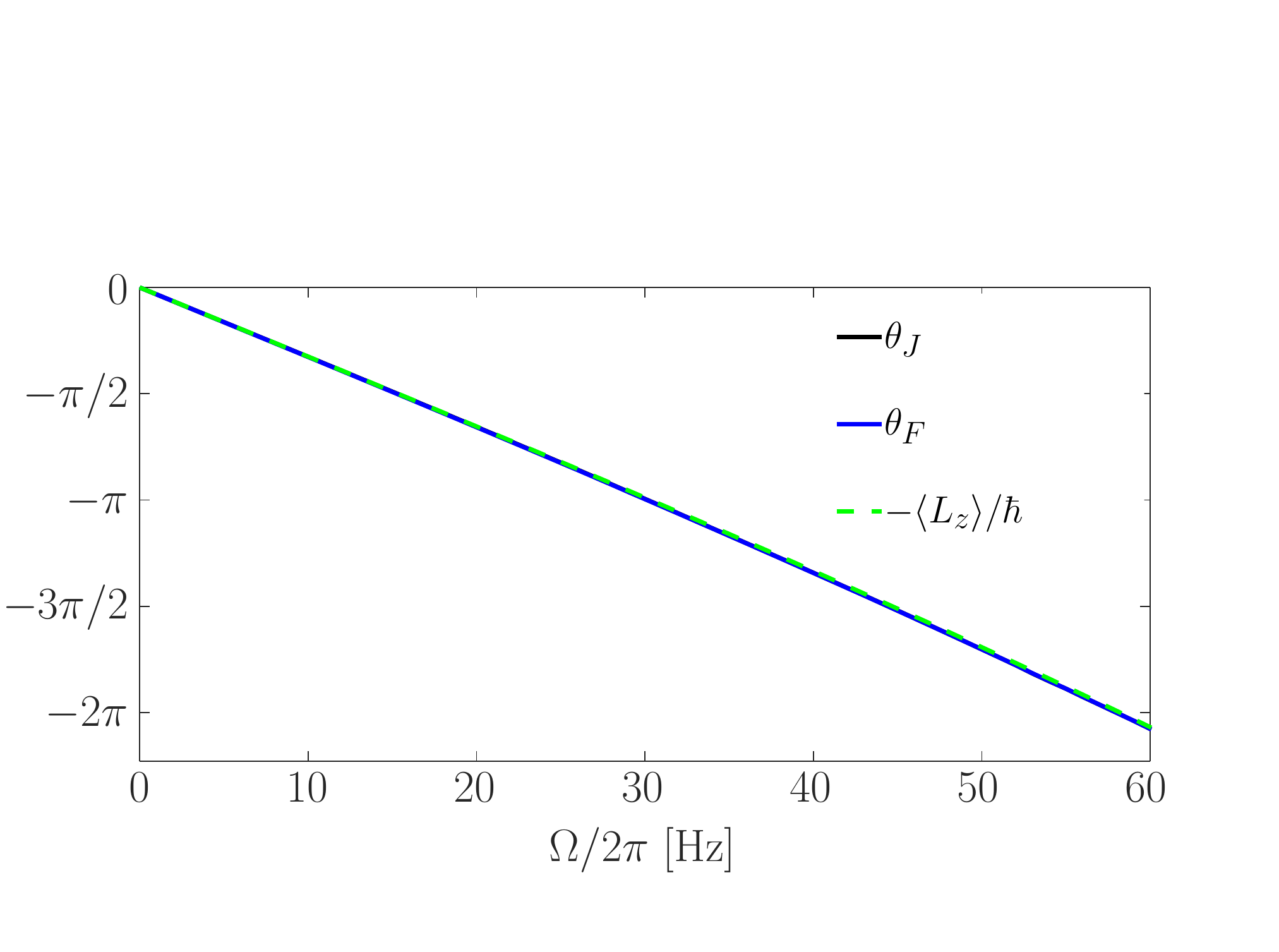}
  \caption{\label{fig:thetas}Phases $\theta _J$ and $\theta _F$ of the
    complex hopping parameters $J$ and $F$ respectively, and
    $-\langle L_z\rangle/\hbar$ as functions of the rotation frequency
    $\Omega$ for the confining potential of the form
    (\ref{eq:trap4}).}
\end{figure}

Let $\mathcal{C}_k(\mathbf{r}_{k,k-1},\mathbf{r}_{k,k+1})$ be the
circulation through the localized function $w_k$, from the junction
$\mathbf{r}_{k,k-1}$ to the junction $\mathbf{r}_{k,k+1}$, and
$ \Delta \beta_k(\mathbf{r}_{k,k-1})$ be the jump of the phase in the
junction between sites $k$ and $k-1$ produced by the imprinted
velocity. The coordinates $\mathbf{r}_{k,k\pm 1}$ mark the positions
of the junctions between the sites $k$ and $k\pm 1$.  Using the
time-dependent multimode model variables $\varphi_k(t)$, the
circulation $\mathcal{C}_k$ must satisfy
\begin{equation}
  \sum_k \mathcal{C}_k(\mathbf{r}_{k,k-1},\mathbf{r}_{k,k+1})
  \frac{m}{\hbar} + \sum_k \varphi_k(t) 
  + \sum_k \Delta
  \beta_k(\mathbf{r}_{k,k-1}) = 2\pi \, l(t)
\label{circ0}
\end{equation}
where $l(t)$ is related to the particular dynamics and is defined by
$ \sum_k \varphi_k(t) = 2\pi \, l(t) $, where in this case we take
$ |\varphi_k(t )| < \pi $ to correctly define the direction of the
associated time-dependent velocity field in the junctions.  Then, we
obtain
\begin{equation}
  \sum_k\mathcal{C}_k(\mathbf{r}_{k,k-1},\mathbf{r}_{k,k+1}) \frac{m}{\hbar}  +  \sum_k \Delta \beta_k (\mathbf{r}_{k,k-1}) =  0 .
\label{circ1}
\end{equation}
From the symmetry of the lattice, the jump in the phase is
$\Delta\beta_k=\Delta\beta=\Theta$.  If the velocity field in the
localized WL function $w_k$ is homogeneous, and hence
$\mathbf{V}^k(\mathbf{r})=\mathbf{\Omega}\times\mathbf{r}_{\mathrm{cm}}^k$,
one can use  (\ref{eq:WLCM}) to calculate the circulation from
$\mathbf{r}_{k,k-1}$ to $\mathbf{r}_{k,k+1}$ as their phase
difference, yielding
\begin{equation}
  \mathcal{C}^H_k=    ( \mathbf{r}_{k,k+1}  -  \mathbf{r}_{k,k-1})  \cdot  (\mathbf{ \Omega} \times \mathbf{r}^k_{\mathrm{cm}}).
\label{circ2}
\end{equation}
Taking into account the lattice symmetry, we have
$|\mathbf{r}_{k,k+1}|=|\mathbf{r}_{k,k-1}|$ and $|\mathbf{r}^k_{\mathrm{cm}}|=|\mathbf{r}_{\mathrm{cm}}|, \forall
k$. Additionally, from geometric considerations for circularly
symmetric WL functions we can further simplify (\ref{circ2}) to obtain
\begin{equation}
\mathcal{C}^H_k = \Omega  |\mathbf{r}_{\mathrm{cm}}|^2 \sin(  2 \pi  / N_c)
\end{equation}
and hence, in terms of the average angular momentum
\begin{equation}
\Theta^{H} = -  \frac{\langle L_z\rangle}{\hbar} \,   \sin(  2 \pi  / N_c).
\label{circ}
\end{equation}
In summary, for isotropic localized densities there exist a clear
correspondence between the shift $\Theta$ and the angular momentum
per particle given by  (\ref{circ}). We numerically calculated
$\langle L_z\rangle / \hbar$, and the phases $\theta _J$ and
$\theta _F$ according to (\ref{J}) and (\ref{F}) for the
potential trap (\ref{eq:trap4}) with $N_c=4$ by calculating the
hopping parameters defined in (\ref{Jjk}) and (\ref{R}),
respectively.  The results are shown in figure \ref{fig:thetas}. The
calculation confirms that there is a unique common shift $\Theta$ for
the two hopping parameters and that its value follows the angular
momentum as predicted by (\ref{circ}) when the velocity profile is
homogeneous. Moreover, a slightly nonlinear dependence of $\Theta$
with $\Omega$ can be observed in figure \ \ref{fig:thetas} and can be
attributed to the increase of $|\mathbf{r}_{\mathrm{cm}}|$ with rotation. We recall
$\langle L_z\rangle$ given by (\ref{Lzgeneral}) with
$\epsilon_0=0$ yields $\langle L_z\rangle =m \Omega |\mathbf{r}_{\mathrm{cm}}|^2$.

\subsubsection{Relation between $\Theta$ and the Peierls phase }

The rotation of a system at an angular frequency $\Omega$ gives rise
to the effective vector potential
$ \mathbf{A}(\mathbf{r})= \mathbf{ \Omega} \times \mathbf{r}$ whose
circulation around a lattice plaquette determines the so-called
Peierls phase \cite{gold14}.  Given that for
$\mathbf{\Omega}=\Omega\hat{z}$,
$ \nabla \times \mathbf{A}(\mathbf{r})= 2 \Omega \hat{z}$ one can
calculate such circulation around a given closed curve using the
Stokes theorem as,
\begin{equation}
\oint  \mathbf{A}(\mathbf{r}).  d\mathbf{r} =    N_c  \,  2 \Omega  \,  S,
\label{flux}
\end{equation}
where $ N_c S$ is the area enclosed by the curve. Contrary to the
derivation using the vector potential, the calculation of the velocity
field circulation (\ref{circ2}) is independent of the curve and it
only depends on the positions of the junctions. Moreover, we note that
as our sites contain many particles the potential minima in general do
not coincide with the centers of mass of the densities
$\mathbf{r}_{\mathrm{cm}}^k$ defined by the WL functions in the $k$
sites.  However, we show below that if one defines the closed polygon
with vertices in each center of mass $\mathbf{r}_{\mathrm{cm}}^k$ and in
the positions of the junctions $\mathbf{r}_{k,k+1}$ and
$\mathbf{r}_{k,k-1}$, the Peierls phase and our result derived from
the homogeneous velocity field coincide.  Given that
$ (\mathbf{r}_{k,k+1} - \mathbf{r}_{k,k-1}) \perp
\mathbf{r}^k_{\mathrm{cm}}$, the points $\mathbf{r}_{k,k+1}$,
$ \mathbf{r}_{k,k-1}$, $ \mathbf{r}^k_{\mathrm{cm}}$, and the origin of
coordinates form a kite whose area is
$ S=|\mathbf{r}_{k,k+1} - \mathbf{r}_{k,k-1} | |\mathbf{r}^k_{cm}|/2
$.  Hence, the circulation of the velocity field along the $k$ site
given by (\ref{circ2}) may be rewritten as
\begin{equation}
 \mathcal{C}^H_k=    ( \mathbf{r}_{k,k+1} -  \mathbf{r}_{k,k-1} )  \cdot
 (\mathbf{ \Omega} \times \mathbf{r}^k_{\mathrm{cm}}) =  2 \Omega S.
\label{circ2b}
\end{equation}
Such a result gives a total circulation $ 2 \Omega S N_c$ in
accordance with Peierls phases.  The positions of the center of mass
$\mathbf{r}_{\mathrm{cm}}^k$ and of the junctions $\mathbf{r}_{k,k+1}$
for each localized WL function are defined by
\begin{align}
 \mathbf{r}_{\mathrm{cm}}^k  &=   \int \!\!   d^3r \,  w_k^*(\mathbf{r})    \mathbf{r}\,  w_k (\mathbf{r}),\quad \mathrm{and}
\label{cm} \\
 \mathbf{r}_{k,k+1} &=   \frac{1}{2}\int \!\!   d^2r \,\left[w_k^*(\mathbf{r})+ w_{k+1}^* (\mathbf{r}) \right]\mathbf{r}  
\left[w_k(\mathbf{r})+ w_{k+1} (\mathbf{r}) \right],
\label{jun}
\end{align}
respectively, where the last 2D integral is performed over the plane
that contains $\mathbf{r}_{k,k+1}$ and it is defined by the angle
$ \vartheta = 2 \pi (k+1)/ N_c $.

\begin{figure}
\centering
  \includegraphics[width=0.7\columnwidth]{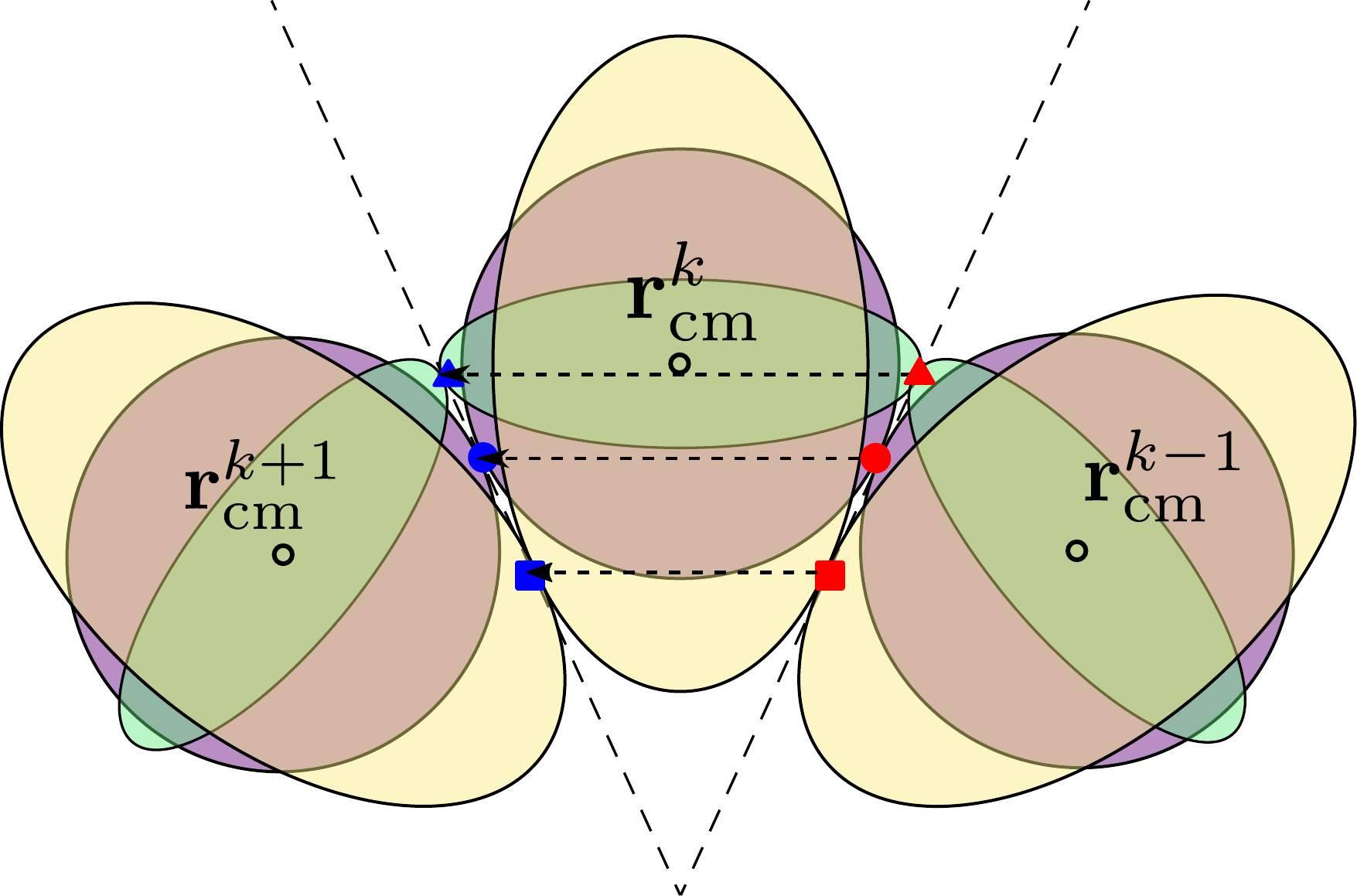}
  \caption{\label{fig:junct}Schematic representation of the WL
    functions (ellipses), their junctions (solid symbols) and their
    center of mass (open  circles) for ring-shaped multiwell
    condensates of different anisotropy. The dashed horizontal arrows
    mark the vectors $\mathbf{r}_{k,k+1}-\mathbf{r}_{k,k-1}$ joining
    the junctions of the $k$-site WL function. The solid circles,
    squares, and triangles correspond to circularly symmetric, prolate
    and oblate WL functions, respectively.}
\end{figure}

The simple correspondence between the angular momentum and $\Theta$
does not exist when the induced velocity field is inhomogeneous.
Still, assuming a homogeneous velocity field, the circulation can be
estimated by  (\ref{circ2}).  As illustrated in
figure \ref{fig:junct}, the anisotropy of the WL functions alters the
position of the junctions yielding a lower circulation for a prolate
WL function, and a higher circulation for an oblate one.

\subsubsection{A numerical example}
\begin{figure}
\centering
  \includegraphics[width=0.7\columnwidth]{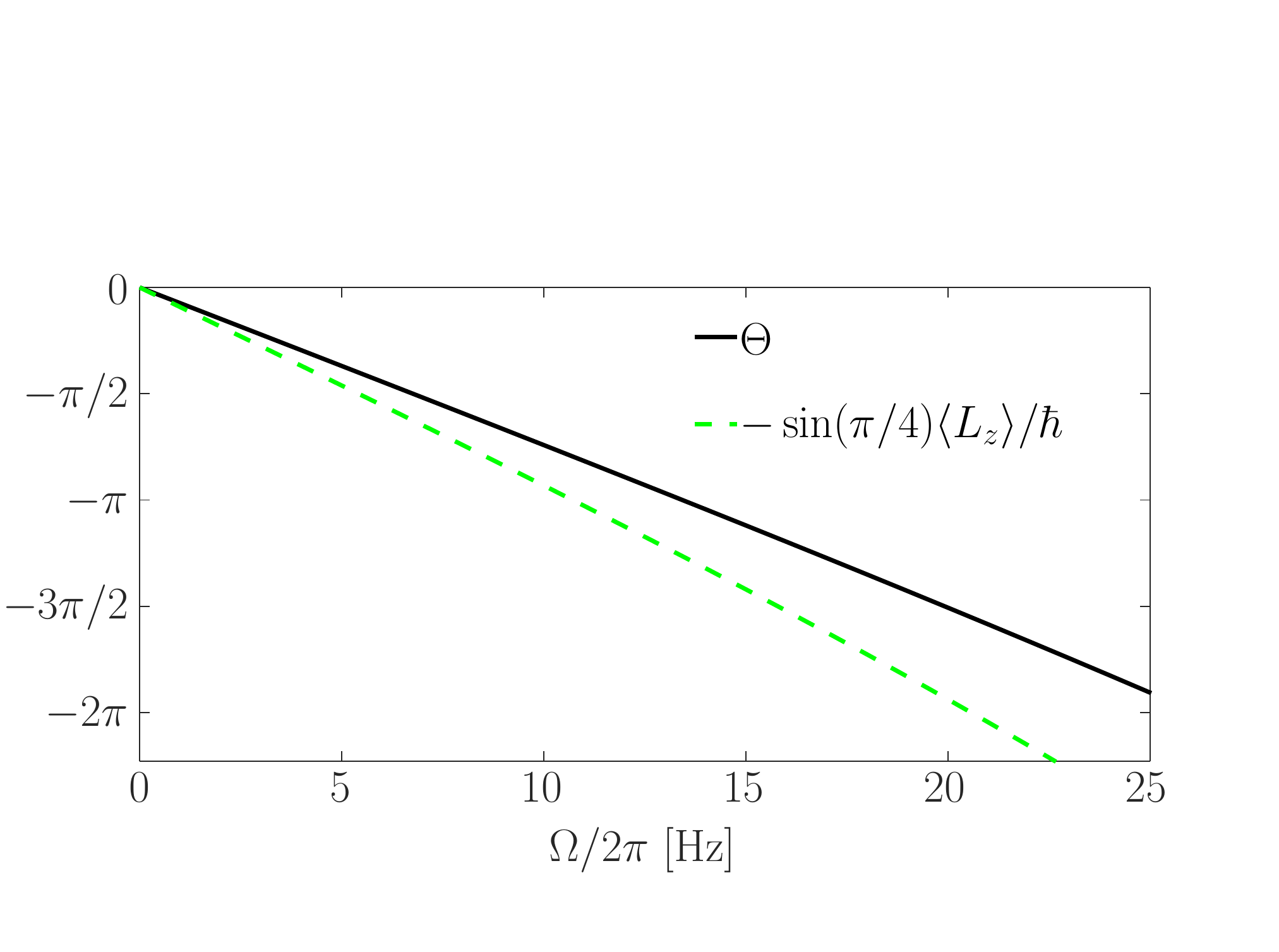}
  \caption{\label{fig:thetas8} Hopping phase
    $\Theta=\theta_J=\theta_F$ and
    $-\sin(\pi/4)\langle L_z\rangle/\hbar$ (cf.  (\ref{circ})) as
    functions of the rotation frequency $\Omega$ for the lattice trap
    potential,  (\ref{eq:latticeTrap}), with $N_c=8$ sites,
    $N=10^4$ , $V_b/\hbar\omega_r=20$, and $\lambda_b/\ell_r=1.6$. }
\end{figure}

The breakdown of the correspondence shown in  (\ref{circ}) between
$\Theta$ and $\langle L_z\rangle$ as a consequence of the anisotropy
of each localized density can be viewed in figure \ref{fig:thetas8} for
the configuration of figure \ref{fig:circu} (d).  The center of mass of
the $k=0$ site is located at
$\mathbf{r}_{\mathrm{cm}}^0 = (4.9385, 2.0456,0)\ell_r$, while the
connecting junctions to the neighboring sites are
$ \mathbf{r}_{0,1} = ( 3.095, 3.095,0)\ell_r$, and
$ \mathbf{r}_{0,-1} = ( 4.3776, 0,0)\ell_r$.  The phase difference
$\mathcal{C}_i^H$ considering an homogeneous velocity field, thus,
yields approximately $0.51$. However, from the intensity of the
absolute value of the velocity field we may see from figure 
\ref{fig:circu} (d), that it decreases from
$ \sqrt{0.024}\hbar/(m\ell_r)$ to $ \sqrt{0.02}\hbar/(m\ell_r)$ and
hence the actual velocity field circulation decreases to approximately
$0.47$, which is in accordance with the value $\Theta = - 0.474$ that
we have obtained from its definition in (\ref{Jjk}).  Such a
correction is due to a nonvanishing eccentricity $\epsilon_0$ and
hence cannot be taken into account by the circulation of the vector
potential $\mathbf{A}$ only.

\subsubsection{\label{sec:junction_speed}The velocity field near the
  junctions}
\begin{figure}
\centering
  \includegraphics[width=0.7\columnwidth]{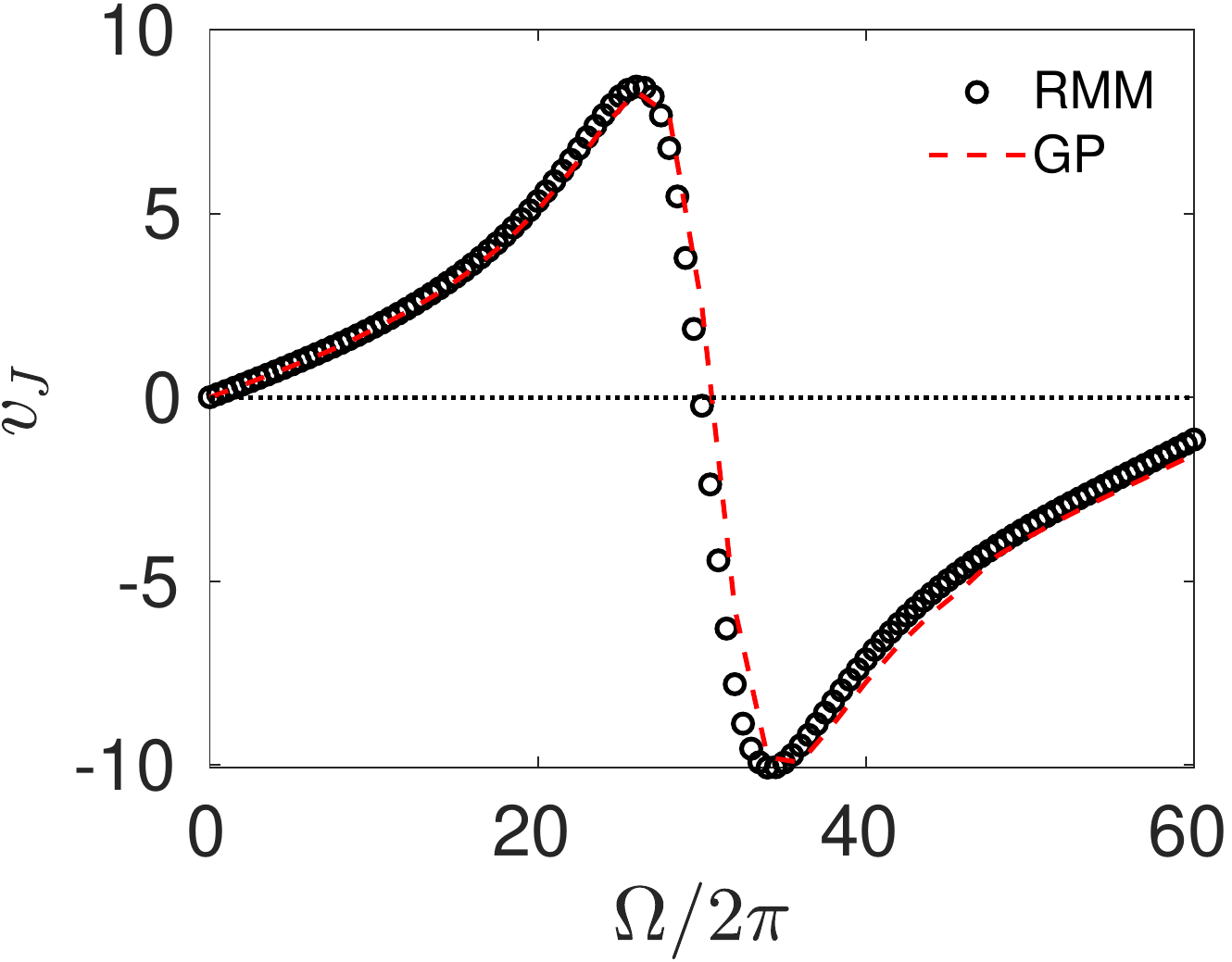}
  \caption{\label{fig:velojunct}Velocity $\mathbf{v}_J=v_J\hat{x}$ at
    the junction $\mathbf{r}_J=(0,y_{\mathrm{CM}},0)$ as function of
    $\Omega$ for the potential trap given by (\ref{eq:trap4}). The
    circles correspond to the RMM model results and the dashed lines
    to the numerical GP solution. The trap parameters are the same as
    those in figure \ref{fig:circu}(c).}
\end{figure}

The correct determination of the phase $\Theta$ within the RMM model
permits an accurate description of the velocity field not only within
the bulk of the onsite localized functions, but also around the
junction. Indeed, in the RMM model the velocity field between
neighbouring sites $k$ and $k+1$ can be obtained with great accuracy
from the order parameter written as a combination of $w_k$ and
$w_{k+1}$ only, as
$\psi_M(\mathbf{r},t)\simeq\sqrt{n_k(t)}e^{i\phi_k(t)}w_k(\mathbf{r})
+ \sqrt{n_{k+1}(t)}e^{i\phi_{k+1}(t)}w_{k+1}(\mathbf{r})$. For the
stationary state with $n=0$, this yields the velocity field
\begin{multline}
  \mathbf{v}(\mathbf{r}) = \frac{1}{N_c\rho}\biggr\{|w_k|^2\mathbf{V}^k(\mathbf{r})
    + |w_{k+1}|^2\mathbf{V}^{k+1}(\mathbf{r}) +   |w_k||w_{k+1}| \cos(\delta\alpha_{k})\left(\mathbf{V}^k(\mathbf{r}) + \mathbf{V}^{k+1}(\mathbf{r})\right)  \\
  \left. + |w_k||w_{k+1}|
    \sin(\delta\alpha_{k})\frac{\hbar}{m}\left(\frac{\nabla
        |w_k|}{|w_k|} - \frac{\nabla |w_{k+1}|}{|w_{k+1}|}\right)
  \right\},
  \label{eq:velojunct}
\end{multline}
where, in the case of the four-site potential well (\ref{eq:trap4}),
one has
$\mathbf{V}^{k}(\mathbf{r})=\mathbf{\Omega}\times
\mathbf{r}_{\mathrm{cm}}^{k}$, and
$\delta\alpha_k =
\dfrac{m}{\hbar}\mathbf{\Omega}\cdot[(\mathbf{r}_{\mathrm{cm}}^{k}-\mathbf{r}_{\mathrm{cm}}^{k+1})\times
\mathbf{r}]$. The velocity field at the point
$\mathbf{r}_J=(0,y_{\mathrm{CM}},0)$ with $y_{\mathrm{CM}}>0$ can be
evaluated from (\ref{eq:velojunct}) taking into account that
$\delta\alpha_k=-\Theta$. In figure \ref{fig:velojunct} we compare
$\mathbf{v}_J=v_J\hat{x}$ obtained from the RMM model and the GP
simulation as a function of $\Omega$.  Given that $\Theta$ depends on
the rotation frequency $\Omega$ as shown in figure \ref{fig:thetas},
the magnitude and sign of $v_J$ is rather sensitive to the rotation.
We may see that the RMM model correctly reproduces the peculiarities
given by the GP equation. In fact, for $\Theta > -\pi$, that
corresponds to frequencies $\Omega/2\pi \lesssim 28$ Hz, the velocity
at the junction is positive, opposing to the $x$ component of the
superposition of the velocity fields coming from the localized states
on neighbouring sites, $\mathbf{V}^{k}(\mathbf{r})$. On the other hand, for
$\Theta<-\pi$, the velocity at the junction reverses and points in the
negative $\hat{x}$ direction.

\section{THE STATIONARY STATES} \label{stationary}

\begin{figure}
\centering
  \includegraphics[width=0.9\columnwidth]{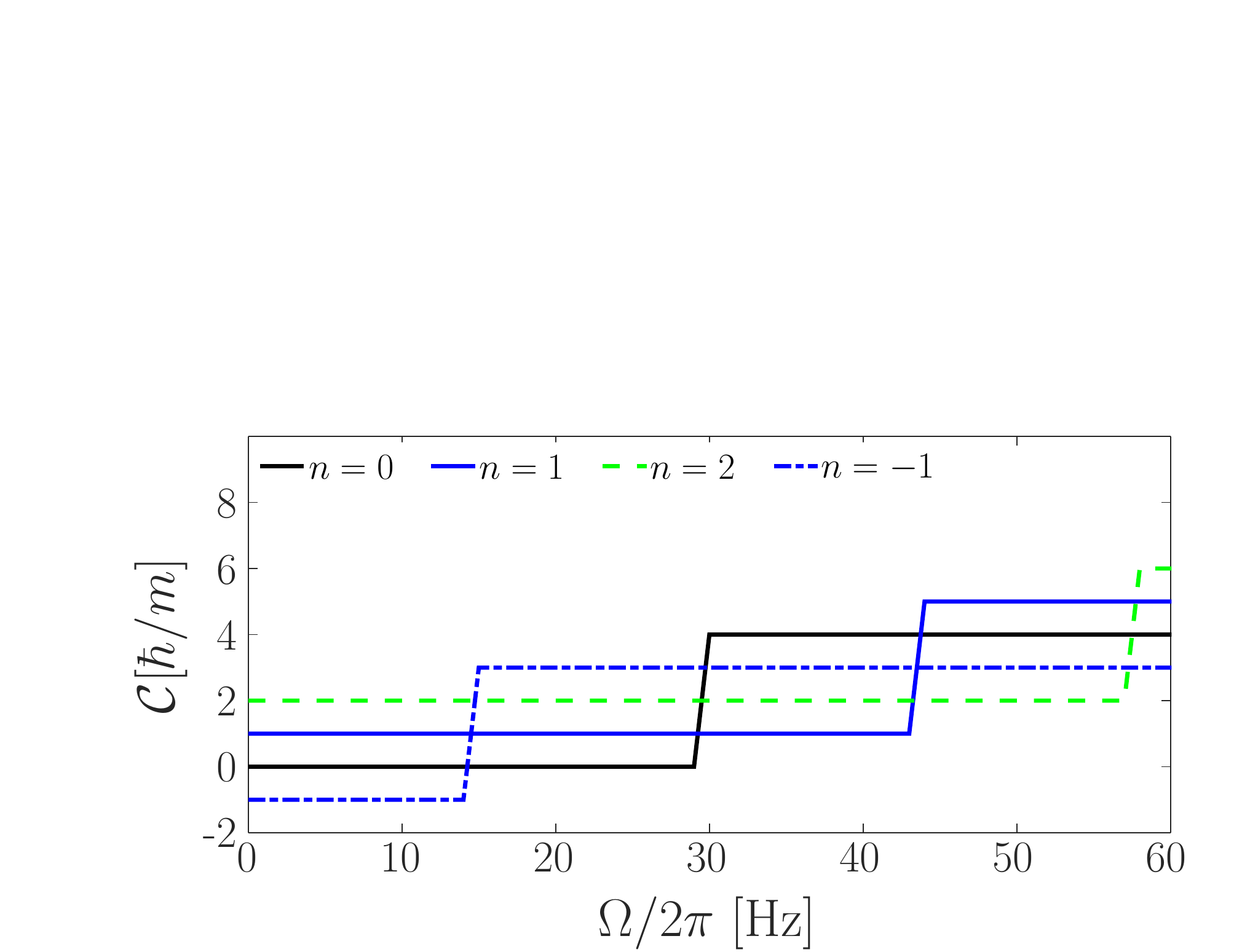}
  \caption{\label{fig:Esrot}Velocity field circulation $\mathcal{C}$
    (in units of $\hbar/m$) for the $n$ stationary state as a function
    of the rotation frequency $\Omega$ for the potential
    (\ref{eq:trap4}). The circulation is calculated along a
    centered square of side $\ell = 2\ell_r$ in the $z=0$ plane. }
\end{figure}

The energy levels $E_n$ can be written in terms of their stationary
states $\psi_n$ as
\begin{equation}
E_n=\int d^3r \,\psi ^*_n\Big [-\frac{\hbar}{2m}\nabla ^2+V_{\mathrm{t}}-\Omega \hat{L}_z+\frac{gN}{2}|\psi _n|^2 \Big ]\psi _n
\label{En}
\end{equation}
where the index $n$ refers to the winding number for $\Omega=0$. Due
to the discrete $N_c$-fold rotational symmetry, the value of the
circulation associated with the stationary state $\psi _n$ can be
equal to the corresponding in a nonrotating case, that is to say
$\hbar /m\, n$, or it can change in amounts of $\hbar /m N_c$. From
figure \ref{fig:Esrot} we numerically confirm this statement by
evaluating the circulation around a centered box of side
$\ell=2\ell_r$ as a function of $\Omega$ for the four stationary
states with lower energy confined by the potential (\ref{eq:trap4}).

In the RMM model the energy levels (\ref{En}) are evaluated by
inverting the basis transformation (\ref{wannier}) and replacing
$\psi_n$ into (\ref{En}). Then, using the definitions for the RMM
model parameters the energy levels take the simple form:
\begin{equation}
E_{n}=\epsilon+\frac{NU}{2N_c}-|K|\cos (\theta _n+\Theta ),
\label{EnNc}
\end{equation}
where $\theta_n=2\pi n /N_c$, and $K=2J+\frac{4}{N_c}F$. It should be
noted that when $4|F|/N_c> 2|J|$, the phase of $K$, $\theta_K$
coincides with $\Theta$, as it happens in our case. Equation
(\ref{EnNc}) provides a useful tool to test the accuracy of the RMM
model. If the potential barriers are not high enough, or if the WL
functions are not properly localized, the energies $E_n$ calculated
from the GP equation will not satisfy (\ref{EnNc}). This can be used
to select the trap parameters for the construction of a reliable
model.

To characterize the energy levels structure it is useful to define the
energy differences $\Delta E_n$ given by
\begin{align}
\Delta E_n&=E_n-E_0+|K|(1-\cos \Theta) \nonumber \\
&= |K|[1-\cos (\theta _n+\Theta )]. 
\label{DeltaEn}
\end{align}
This description emphasizes the band structure of the energy levels in
periodic systems \cite{Perez2007,Ferrando2005}, but with $\theta_n$
shifted in $-\Theta$.

\begin{figure}
  \centering
  \includegraphics[width=0.7\columnwidth]{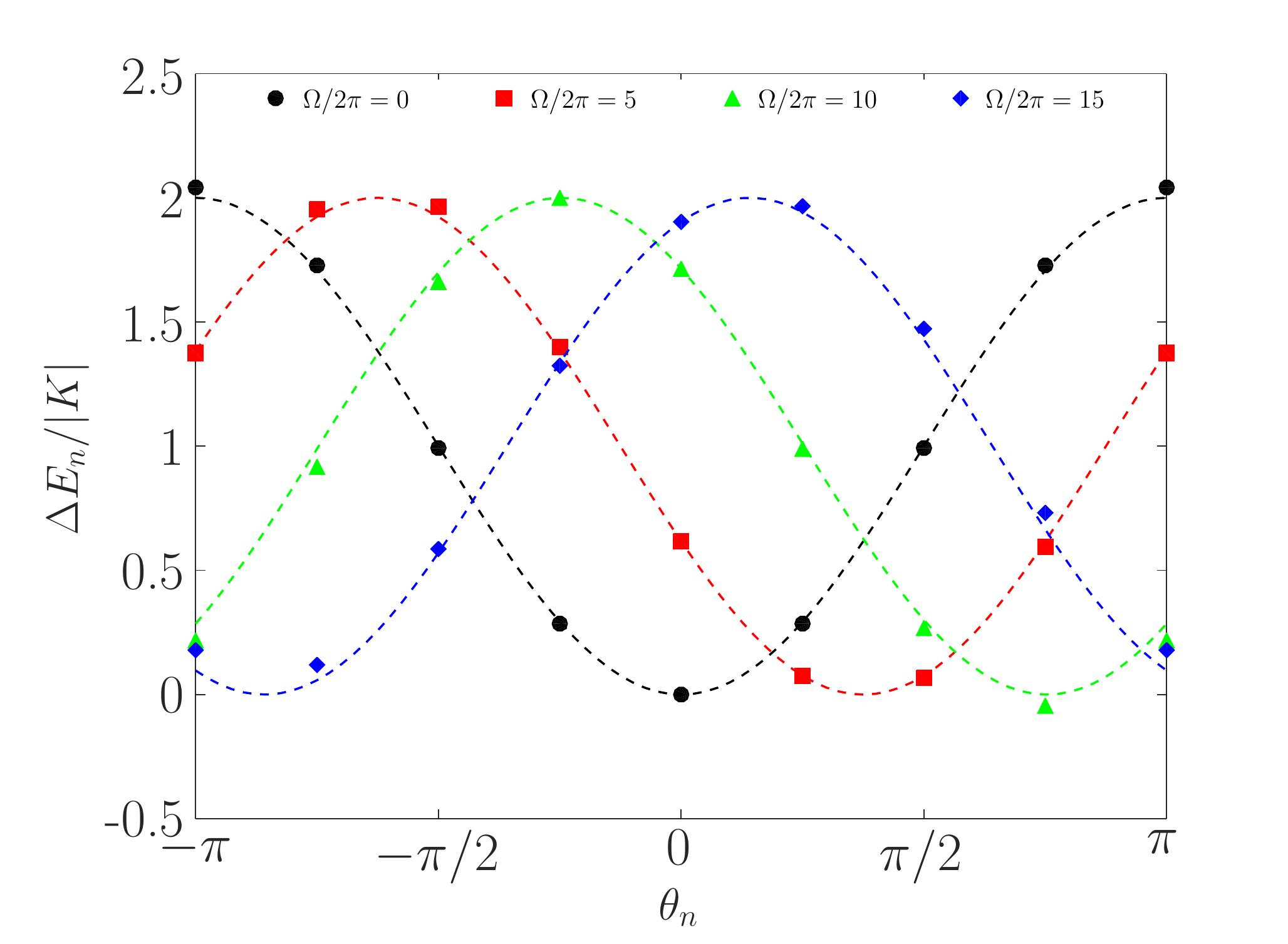}
  \caption{\label{fig:figDO}Band structure of the energy levels $E_n$
    of the lattice trap potential $V_2$ with $N_c=8$ wells for the
    same parameters as in figure \ref{fig:thetas8} and for rotation
    frequencies $\Omega/2\pi=0, 5, 10$, and $15$ Hz corresponding
    $\Theta/\pi=0, -0.37, -0.75$, and $-1.14$, respectively.}
\end{figure}

In figure \ref{fig:figDO} we illustrate the energy structure
(\ref{DeltaEn}) for the case $N_c =8$. The curves correspond to
the second line of the right hand side of (\ref{DeltaEn})
divided by $|K|$ and the symbols are obtained from the energy
differences of the GP stationary states for each $\Omega$. The
agreement between the curves and the symbols confirms the
validity of the RMM model.

The states $\psi _n$ and $\psi _{-n}$, that are degenerate for
$\Omega =0$, reach a maximum energy difference when
$\Theta = -\pi /2$. For this rotation frequency, and assuming even
$N_c$, the states $\psi _{N_c/2}$ and $\psi _0$ become degenerate.
The energy ordering for each $n$ can be easily calculated for
arbitrary $\Omega$ since the entire band moves by an amount
$\Theta(\Omega)$ to the right assuming $\Omega >0$, which in turn is
proportional to the angular momentum in the case of circularly
symmetric onsite densities.  Furthermore, the nonrotating ordering is
restored when $\Theta =2\pi$, which corresponds to the rotation
frequency
\begin{equation}
\Omega _{\mathrm{restore}} =\frac{2\pi \hbar}{m r_{\mathrm{cm}} ^2 \sin (2\pi /N_c)}.
\label{OmegaTheta}
\end{equation}  
This rotation frequency $\Omega_{\mathrm{restore}}$ is further
constrained to being below $\omega_r$ to ensure the confinement of the
WL function.

\section{SPECIAL MULTIMODE DYNAMICS VERSUS GROSS-PITAEVSKII
  SIMULATIONS} \label{nonstationary} We have studied the accuracy of
the RMM model comparing the solutions of (\ref{ncmode1hn}) and
(\ref{ncmode2hn}) with full 3D GP simulations for several initial
conditions and rotation frequencies. In accordance with previous
results for nonrotating traps \cite{Jezek2013,Nigro2018a}, we have
found that to ensure a quantitative agreement between them one ought
to take into account the onsite interaction energy dependence with the
imbalance and hence employ the effective interaction parameter
$U_{\mathrm{eff}}$ instead of the bare onsite $U$. It is worthwhile
noticing that using the RMM model, the running time for the
computation of the time evolution of populations and phase differences
dramatically reduces by more than five orders of magnitude with
respect to the that of a 3D numerical GP simulation.

\begin{figure}
  \centering
  \includegraphics[width=0.45\columnwidth]{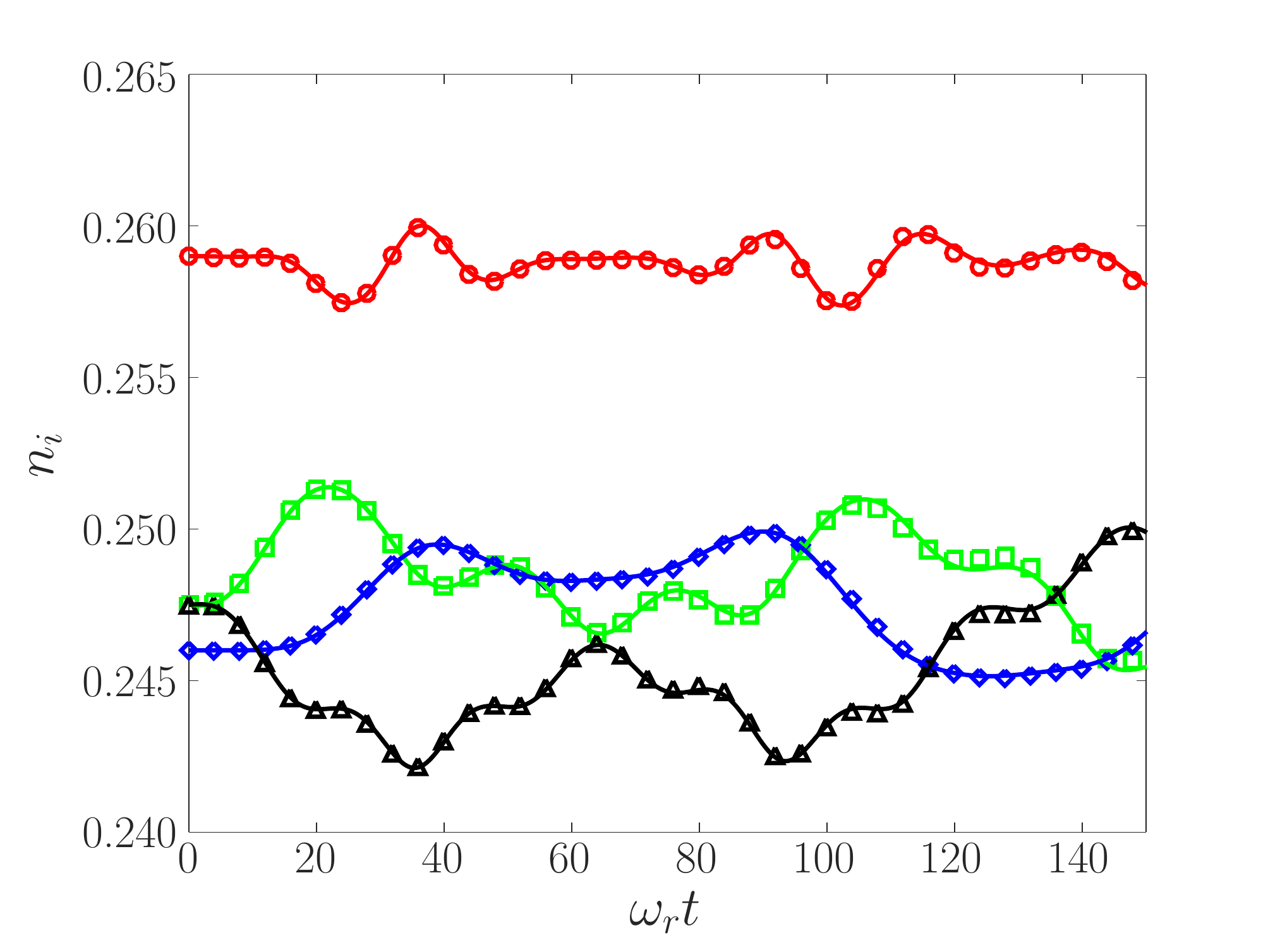}
  \includegraphics[width=0.45\columnwidth]{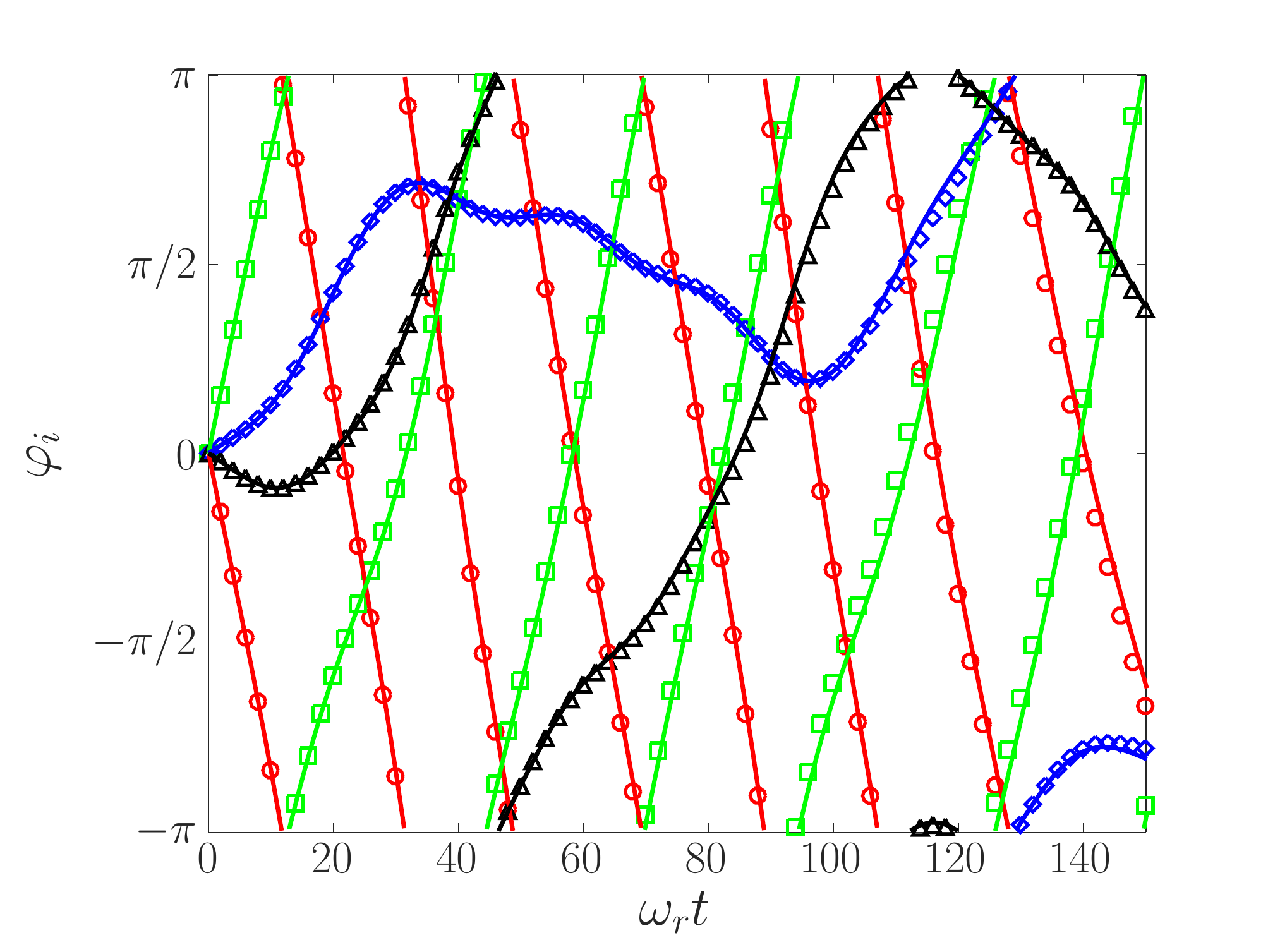}
  \caption{\label{fig:dyn1} Population $n_i$ and phase differences
    $\varphi_i$ as functions of $t$ for $\Omega/2\pi =15.065$Hz. The
    solid lines and symbols correspond to results of the GP equation
    and RMM model, respectively. The initial conditions are given by
    $N_{-1}=2475$ (triangles), $N_0=2590$ (circles), $N_1=2475$
    (squares), and $N_2=2460$ (diamonds) and $\varphi _i=0$. The
    reflection symmetry of the initial condition ($N_{-1}=N_1$) is
    clearly broken.}
\end{figure}

\begin{figure}
  \centering
  \includegraphics[width=0.45\columnwidth]{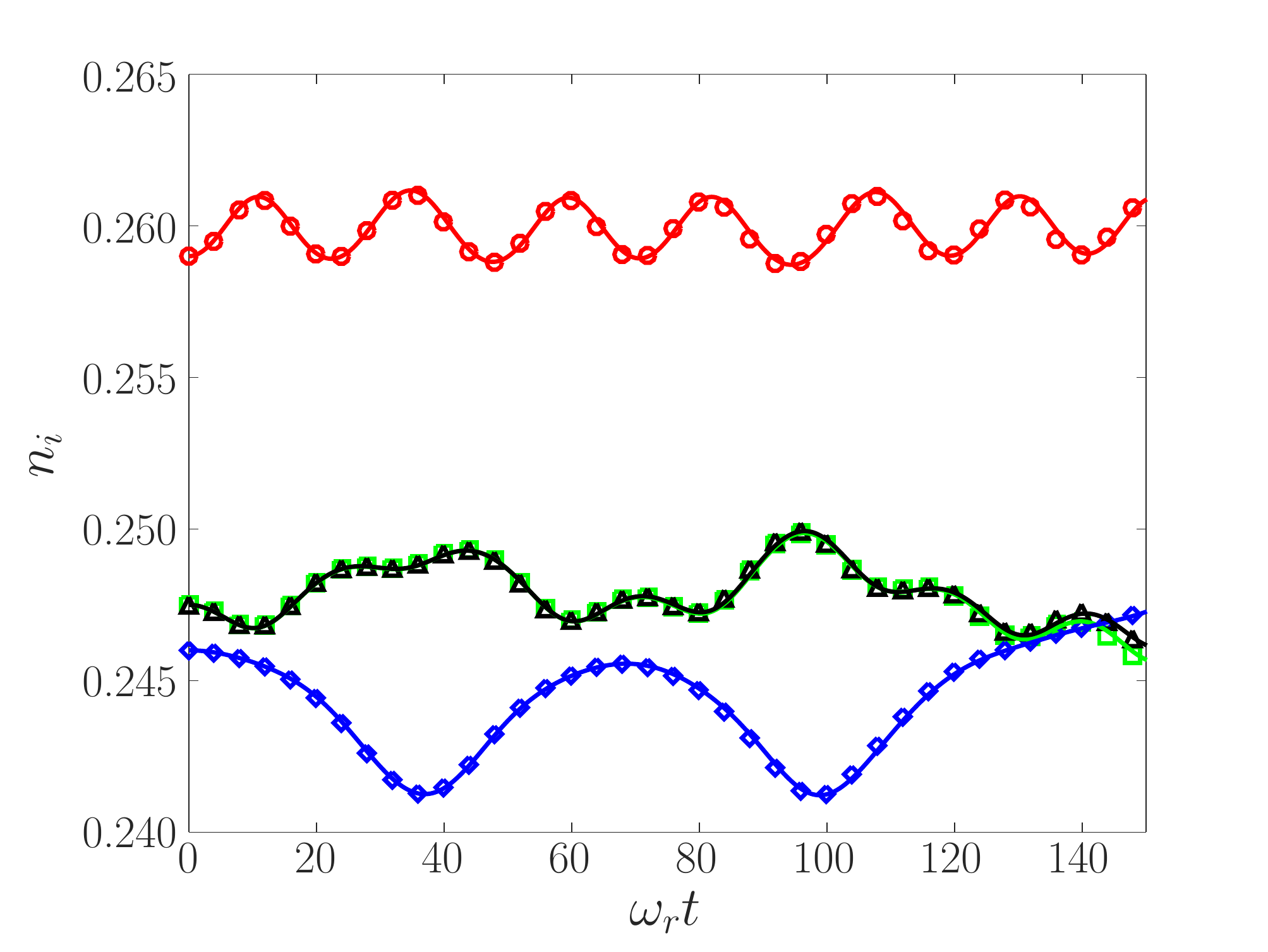}
  \includegraphics[width=0.45\columnwidth]{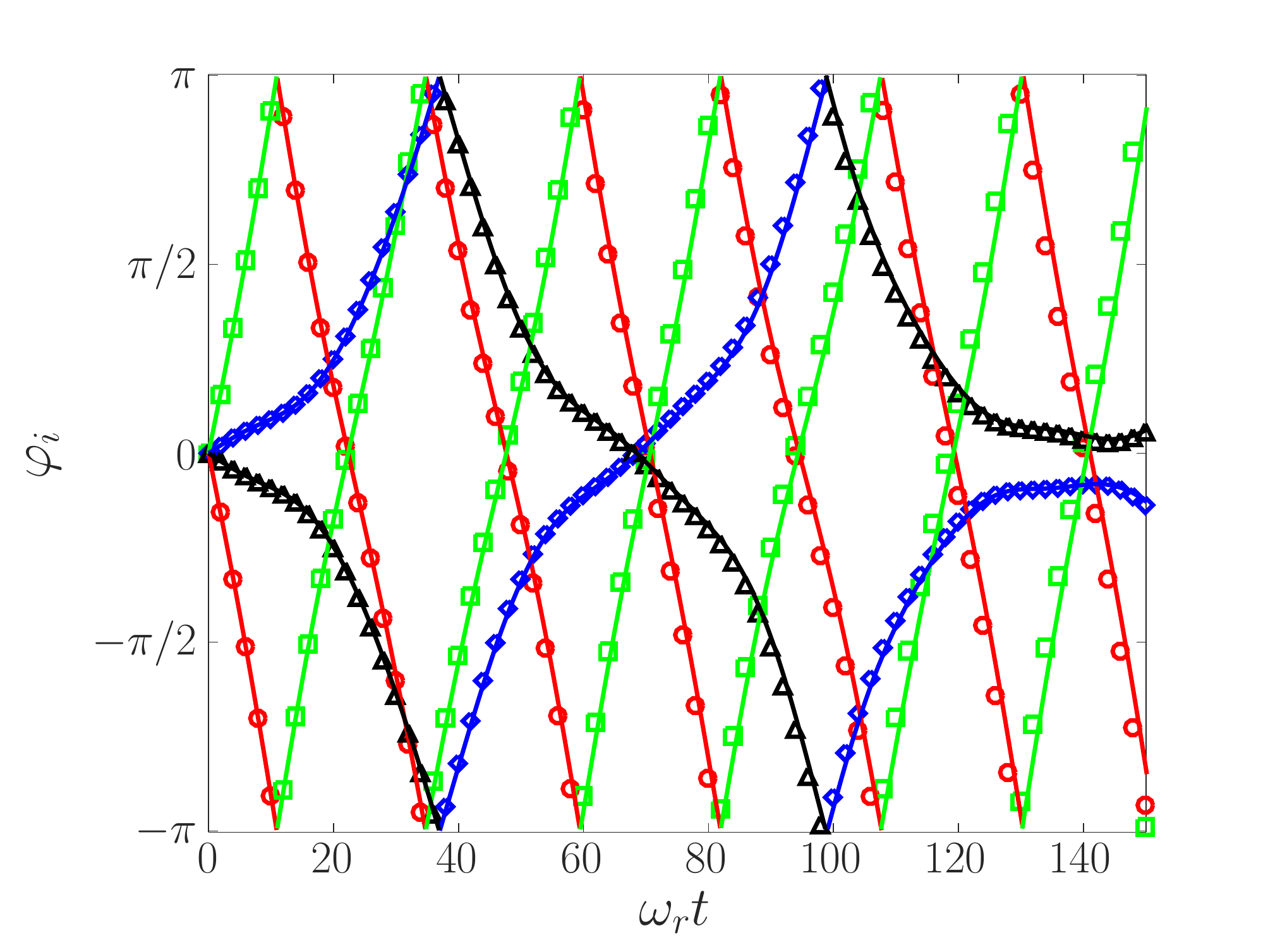}
  \caption{\label{fig:dyn2} Population $n_i$ and phase differences
    $\varphi_i$ as functions of $t$ for $\Omega/2\pi =30.13$Hz for the
    same initial conditions as in figure \ref{fig:dyn1}. The solid lines
    and symbols correspond to results of the GP equation and RMM
    model, respectively. In this case $N_1=N_{-1}$ is maintained
    during the whole evolution.}
\end{figure}

Here we present some results for the four-well trap $V_1$ with
$N=10^4$ and initial conditions given by $N_1=N_{-1}$, $N_0\neq N_2$
and $\varphi _k =0$.  In this case we have found a value of
$\alpha=0.219$ \cite{Nigro2018a}, irrespective of the rotation
frequency.  This choice of initial conditions allows us to analyze a
particular effect of the rotation. If we set $\Omega =0$, the
left-right reflection symmetry of the trapping potential ensures that
$N_1 (t)=N_{-1} (t)$ during the whole evolution. However, rotation
breaks this symmetry in general. In the top (bottom) panel of
figure \ref{fig:dyn1} we show the population dynamics (phase
differences) in each site corresponding to the GP simulations and the
integration of the RMM model (\ref{ncmode1hn}) and
(\ref{ncmode2hn}). We focus on the particular choice of
$\Omega/2\pi = 15.065$Hz, which corresponds to a value of $\Theta$
close to $-\pi/2$ and hence the contributions to $|dn_k/dt|$ and
$|d\varphi_k/dt|$ of the hopping terms are maximum at $t=0$.  It is
clear that in this case the initial symmetry is not maintained, and
that the RMM model appropriately describes the dynamics.
However, the structure of (\ref{ncmode1hn}) and (\ref{ncmode2hn})
allows us to restore this symmetry if $\Theta =l\pi$, being $l$ a
negative integer.  The trap parameters employed in the dynamics with
the potential given by (\ref{eq:trap4}) guarantees a homogeneous
velocity field. Therefore, the associated rotation frequency
$\Omega _{l}$ can be deduced from the mean value of the angular
momentum $L_z=m\Omega\, r_{\mathrm{cm}}^2.$ For $l=-1$ a minus sign
appears in the hopping parameters $J$ and $F$. The $\psi _{\pm n}$
states are degenerate again and the left-right reflection symmetry is
recovered. The population dynamic for this case is shown in the top
panel of figure \ref{fig:dyn2}, where we employed the same initial
condition as in figure \ref{fig:dyn1} but with a rotation frequency
$\Omega =2\pi \times 30.13$Hz which gives a phase $\Theta\simeq
-\pi$.
Again, the RMM model predicts the same dynamics as the GP
simulations. It is worthwhile to notice that such a restoration of the
$N_1(t)=N_{-1}(t)$ symmetry is a pure quantum phenomenon, since
the phase $\Theta$ is associated with the quantization of the velocity field
circulation.

\section{SUMMARY AND CONCLUDING REMARKS} \label{summary}

We have formulated a rotating multimode model for a Bose-Einstein
condensate confined in a ring-shaped optical lattice with $N_c$ sites.
The appearance of induced inhomogeneous phases in the condensate
implies that the onsite localized basis set cannot be taken as real
functions, and hence the multimode hopping parameters $J$ and $F$
become complex numbers with the same phase $\Theta$. This was
confirmed by numerically solving the Gross-Pitaevskii equations for
several trap geometries.

To understand the nature of the induced velocity fields, as a first
step we considered an off-axis rotating single condensate confined by
an anisotropic harmonic trap. Varying the trapping frequencies in the
orthogonal direction to the axis of rotation, we observed that the
induced velocity field tilts in the direction of growth of the
velocity modulus, and this corresponds to a density profile that has
no axial symmetry with respect to its center of mass. When this axial
symmetry is restored, the velocity field becomes homogeneous.  On this
last case, the complex phases of the localized basis can be easily
predicted and a simple analytical relation between the hopping phase
and the angular momentum is found by calculating the velocity field
circulation.  We have shown how these nontrivial imprinted phases can
be analytically understood using the continuity equation for the
density in the rotating frame.

In a second step, the induced velocity field was studied in rotating
ring-shaped optical lattices for several geometries. It was found that
for lattices in the tight-binding regime the velocity fields can be
described in the same manner as in the single condensate case.  For
onsite homogeneous velocity fields, the phases of the hopping
parameters are consistent with the Peierls phases appearing in systems
subject to effective vector potentials. On the other hand, the effect
of an inhomogeneity in the velocity field due to the lack of circular
symmetry of the localized densities cannot be accounted for by the
Peierls substitution formula alone. The full definition of the hopping
parameters must be used to correctly calculate their phases. Finally,
the validity of the rotating multimode model was verified for the
first time by comparing its predictions with those obtained by
numerically integrating the Gross-Pitaevskii equation for several
initial conditions. In particular, we tested the rotating multimode
model against nontrivial symmetry-preserved initial conditions and
found they are accurately reproduced by the model. Finally, the RMM
constitutes an extremely fast and accurate tool to predict the
evolution of the population and phase differences, allowing to tackle
also the dynamics of the velocity fields in multiple well systems. The
model thus provides a promising tool to investigate features of the
more complicated vortex dynamics. Work in this direction is in
progress.

\ack
 This work was supported by CONICET and Universidad de Buenos
 Aires through grants PIP 11220150100442CO and UBACyT
 20020150100157,
 respectively.

 \appendix
 \section*{Appendix: Selection of the lattice parameters: balance of the energy
   contributions}
 \setcounter{section}{1}
 In this Appendix we show how the confining potential of a ring-shaped
 lattice of the form (\ref{eq:latticeTrap}) can be constructed in
 order to obtain an almost uniform velocity field in each site. This
 is analyzed by comparing the most important contributions to the
 energy for varying parameters of the confinement.
 
 If one considers small barrier widths, one can assume that
 $\nabla \rho $ lies in the $\hat{r}$ direction far from the potential
 barriers, and hence the velocity field in the bulk can be
 approximated by $\mathbf{V} (\mathbf{r})= \dfrac{\mathcal{A}}{r} \hat{\theta}$,
 which verifies the continuity equation,
 $\nabla .  [\rho ( \mathbf{V}(\mathbf{r})- \mathbf{ \Omega} \times
 \mathbf{r})] = 0$. The amplitude $\mathcal{A}$ can be later chosen by enforcing
 a particular value of the angular momentum.  On the other hand, in a
 general rotating optical lattice the velocity profile is determined
 by the competition between the increase of the kinetic energy due to
 the phase gradient in the bulk, and the reduction of the angular
 momentum term in the energy. We shall call such an energy balance
 $E_r$, which can be analyzed in the first quadrant thanks to the
 discrete symmetry of the lattice. Hence, writing the WL function at
 the first site ($k=0$) as
 $w_0(\mathbf{r})=|w_0(\mathbf{r})|e^{i\alpha_0(\mathbf{r})}$, this
 energy is given by
\begin{equation}
E_r \simeq \int d^3r\, w^*_0(\mathbf{r})\bigg (\frac{\hbar ^2}{2m}|\nabla \alpha _0 (\mathbf{r})|^2-\Omega \hat{L}_z \bigg )w _0(\mathbf{r}).
\label{Er}
\end{equation}
Although (\ref{Er}) cannot be analytically computed in general, an
expression in terms of the mean values of the angular momentum per
particle and of the spatial coordinates can be obtained in two special
limits: a) when the onsite localized density is circularly symmetric,
and b) when the barrier widths are small enough and thus $\nabla\rho$
lies in the radial direction. In a four-well trap ($N_c=4$), the WL
function in the first site can be written as
\begin{align}
  w^a_0 (\mathbf{r}) &=|w^a_0(\mathbf{r})|e^{\textstyle i\frac{\langle \hat{L}_z \rangle}{\hbar}\frac{y-x}{\langle x\rangle +\langle y\rangle}} \label{wa0} 
\end{align}
for case a), and
\begin{align}
w^b_0 (\mathbf{r}) &= |w^b_0 (\mathbf{r})|e^{\textstyle i\frac{\langle \hat{L}_z \rangle}{\hbar}\theta} \label{wb0}
\end{align}
for case b).  Inserting (\ref{wa0}) and (\ref{wb0}) into
(\ref{Er}) we obtain
\begin{eqnarray}
&& E^{a}_r=\frac{1}{2m}\langle \hat{L}_z \rangle ^2\frac{1}{\langle x\rangle ^2+\langle y\rangle^2}-\Omega \langle\hat{L}_z\rangle \label{Era} \\ 
&& E^b_r=\frac{1}{2m}\langle\hat{L}_z\rangle^2\langle\frac{1}{x^2+y^2}\rangle-\Omega \langle\hat{L}_z\rangle \label{Erb} \\  
&& \nonumber
\end{eqnarray} 
for case a) and b), respectively.  Therefore, the sign of
$\Delta E_r=E^b_r-E^a_r$ indicates the energetically favored velocity
field for the system. For $\Delta E_r <0$, curved velocity fields are
favored; whereas for large $\Delta E_r >0$ the velocity profiles are
expected to be linear in each site and, in particular, consistent with
(\ref{eq:WLCM}). To confirm this connection between the
minimization of $E_r$ and the velocity field curvature we numerically
studied the velocity dispersion
$\sigma _v^2 =\langle\mathbf{v}^2\rangle-\langle\mathbf{v}\rangle^2$
in a given site for a set of values $(\lambda _b ,V_b)$ corresponding
to the potential $V_2$ (cf. (\ref{eq:latticeTrap})).

\begin{figure}
  \centering
  \includegraphics[width=0.7\columnwidth]{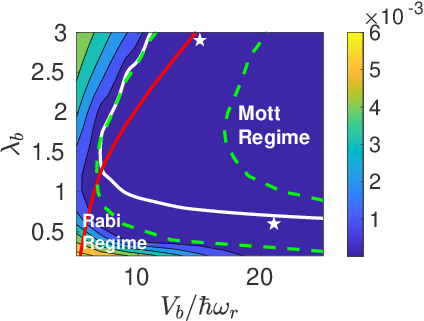}
  \caption{\label{fig:sigmav} Velocity dispersion $\sigma_v$ (in
    arb. units) as a function of the lattice parameters $V_b$ and
    $\lambda_b$. The white solid curve marks $\Delta E_r =0$, while
    the red solid one marks the curve defined by
    $V_b=\mu(\Omega=0,\lambda_b,V_b)$. The region inside the dashed
    lines corresponds to the tight-binding regime bounded by Rabi
    (from below) and Mott-insulator (from above) regions. The white
    stars correspond to the values of $(\lambda_b, V_b)$ in the top
    panels of figure \ref{fig:circu}.}
\end{figure}

In figure \ref{fig:sigmav} we present results for the dispersion
$\sigma_v$ in the lattice potential $V_2$ with $N_c=4$.  We show in
colors the value of $\sigma _v$ in the $(\lambda _b , V_b)$ parameter
space for $\Omega \simeq 0.014\omega_r$. The homogeneous velocity
field region should lie to the right of the red solid curve marking
where the minimum of the barrier equals the chemical potential in the
absence of rotation, which is the condition for the condensates to be
weakly linked. Below (above) the white curve, for which
$\Delta E_r =0$, we have that $\Delta E_r<0$ ($\Delta E_r >0$). At the
region where $\Delta E_r =0$, the velocity dispersion begin to
decrease asymptotically to zero. This behavior confirms that the
velocity field induced by the rotation of the lattice can be
associated with the minimization of $E_r$.


\begin{thebibliography}{99}
\bibitem{note1} {In a rotating harmonic trap, due to the centrifugal
    force the center of mass, which coincides with the center of
    symmetry, is shifted to
    $x_{cm}=x_0 \omega_x^2/(\omega_x^2-\Omega^2)$ with respect to the
    center of the harmonic trap $\mathbf{r}_0=(x_0,0,0)$.}
\end{thebibliography}
\end{document}